\documentclass[%
reprint,
superscriptaddress,
frontmatterverbose,
preprintnumbers,
longbibliography,
amsmath,amssymb,
aps,
pra,
floatfix,
notitlepage,
nofootinbib,
]{revtex4-2}

\usepackage{lipsum}

\usepackage{graphicx}
\usepackage{dcolumn}
\usepackage{bm}
\usepackage{hyperref}
\usepackage{xspace}
\usepackage{subfigure}
\hypersetup{
colorlinks=true,
hypertexnames=false,
linkcolor=blue,
filecolor=blue,
citecolor=blue,
urlcolor=blue,
}

\newcommand{\mycomment}[1]{}

\usepackage{comment}

\usepackage{thmtools}
\usepackage{thm-restate}

\usepackage{enumerate}
\usepackage{enumitem}

\usepackage{amssymb}
\usepackage{mathtools}

\usepackage{mathrsfs}
\usepackage{multirow}
\usepackage{bbm}

\usepackage[dvipsnames]{xcolor}

\definecolor{sanddune}{rgb}{0.59, 0.44, 0.09}
\definecolor{darkblue}{RGB}{0,0,102}
\definecolor{darkred}{rgb}{0.5,0.,0.}
\definecolor{BlueViolet}{RGB}{138,43,226}
\definecolor{SkyBlue}{RGB}{30,144,255}
\definecolor{DarkGreen}{RGB}{0,100,0}

\usepackage{amsthm}
\usepackage{amsmath}
\theoremstyle{plain}
\newtheorem{thm}{Theorem}
\newtheorem{lem}[thm]{Lemma}
\newtheorem{prop}[thm]{Proposition}

\theoremstyle{definition}
\newtheorem{defn}{Definition}

\newcommand{\abs}[1]{\left|#1\right|}

\newcommand{\mc}{\mathcal}

\newcommand{\mbb}{\mathbb}

\newcommand{\supp}{\operatorname{supp}}
\newcommand{\w}{\mathrm{W}}
\newcommand{\wopt}{\mathrm{W}_{\!\mathrm{opt}}}
\newcommand{\wavg}{\mathrm{W}^{\mathrm{avg}}}
\newcommand{\woptavg}{\mathrm{W}_{\!\mathrm{opt}}^{\mathrm{avg}}}
\newcommand{\wlb}{\mathrm{W}_{\!\mathrm{LB}}}
\newcommand{\comments}[1]{} 

\newcommand{\CC}[1]{\ensuremath{\mathsf{#1}}\xspace}
\newcommand{\NP}{\CC{NP}}
\newcommand{\NPcomplete}{\ensuremath{\NP\text{-complete}}\xspace}
\newcommand{\NPhard}{\ensuremath{\NP\text{-hard}}\xspace}

\newcommand{\Prob}[1]{\textsc{#1}\xspace}
\newcommand{\MW}{\Prob{MW-SG}}
\newcommand{\MLD}{\Prob{MLD}}
\newcommand{\SBP}{\Prob{SBP}}
\newcommand{\SBPFtwo}{\SBP\ensuremath{_{\mathbb{F}_2}}\xspace}
\newcommand{\ThreeDM}{\Prob{3-Dimensional Matching}}
\newcommand{\leqp}{\le_p}

\newcommand{\ba}{\begin{eqnarray}}
\newcommand{\ea}{\end{eqnarray}}

\DeclareMathOperator{\Tr}{Tr}

\newcommand{\wt}{\mathrm{wt}}

\newcommand{\x}{\mathbf{x}}

\newcommand{\s}{\mathbf{s}}
\newcommand{\e}{\mathbf{e}}

\usepackage{makecell}
\usepackage{multirow}

\usepackage{autobreak}

\usepackage{algorithm}
\usepackage{algpseudocode}

\begin{document}
\newcommand{\onenorm}[1]{\left\| #1 \right\|_1}
\newcommand{\twonorm}[1]{\left\| #1 \right\|_2}

\title{Theory of low-weight quantum codes}

\author{Fuchuan Wei}
\affiliation{Yau Mathematical Sciences Center, Tsinghua University, Beijing 100084, China}

\author{Zhengyi Han}
\affiliation{Yau Mathematical Sciences Center, Tsinghua University, Beijing 100084, China}

\author{Austin Yubo He}
\affiliation{Yau Mathematical Sciences Center, Tsinghua University, Beijing 100084, China}

\author{Zimu Li}
\affiliation{Yau Mathematical Sciences Center, Tsinghua University, Beijing 100084, China}

\author{Zi-Wen Liu}
\affiliation{Yau Mathematical Sciences Center, Tsinghua University, Beijing 100084, China}


\begin{abstract}
Low check weight is a  crucial code property for fault-tolerant quantum computing, which underlies the strong interest in quantum low-density parity-check (qLDPC) codes.
Here, we initiate the theory of weight-constrained stabilizer codes from various foundational perspectives including the complexity of computing code weight and the explicit boundary of feasible low-weight codes in both theoretical and practical settings.
We first prove that computing the optimal generator weight of a stabilizer code is $\mathsf{NP}$-hard, motivating efficiently computable bounds.
We derive analytical lower bounds on check weight in terms of code rate and distance, identifying the minimum weights needed for single-qubit error detection and correction, as well as the sharp distance and rate limits of weight-three error-detecting codes.
Matching constructions show that these bounds are tight in several regimes.
To establish refined finite-size constraints, we develop a linear programming framework based on quantum weight enumerators subject to generator-weight constraints, yielding exact optimal weights for all parameter combinations with $n\le9$. Finally, we show that the same framework can incorporate architecture-dependent constraints, using the 127-qubit IBM Eagle chip as a concrete example.
Our study brings the weight as a crucial parameter into coding theory and provides guidance for code design and utility in practical scenarios.
\end{abstract}

\pacs{}
\maketitle


\section{Introduction}

The transformative promise of quantum computing and various other technologies~\cite{Nielsen_Chuang_2010,Shor_1997,Lloyd:1996aai,Giovannetti:2011chh} is tempered by a fundamental practical challenge: quantum systems and their manipulation are inherently susceptible to  noise and errors, necessitating efficient fault tolerance schemes in order to realize scalable advantages in practice.

Such fault tolerance is expected to be built on quantum error-correcting (QEC) codes~\cite{shor1995qec,gottesman1997stabilizer,Shor1996FaultTolerant,Campbell_2017}, which protect logical quantum information by encoding it into a suitable larger Hilbert space, enabling physical errors to be detected and corrected.
Stabilizer codes provide a canonical framework for QEC in which error correction is enabled by measuring certain parity-check (stabilizer) operators and applying recovery operations accordingly~\cite{gottesman1997stabilizer}.
While the code rate and distance are regarded as the most fundamental code parameters, check weight (i.e., the size of the nontrivial support of stabilizers) is a less studied property that is nonetheless particularly crucial in the fault tolerance context, as higher-weight checks generally require larger circuits and more ancilla resources, worsening experimental complexity and error propagation.

The importance of low check weight has driven numerous research directions of substantial interest, including quantum low-density parity-check (qLDPC) codes~\cite{10.5555/2685179.2685184, breuckmann_qldpc_2021}, generalized coding schemes such as subsystem codes~\cite{Poulin_2005,Aliferis07:subsystem} and dynamical codes~\cite{Hastings2021dynamically,fu2024errorcorrectiondynamicalcodes}, and weight reduction methods~\cite{hastings_weight_2016,hastings2021quantum,sabo_weight-reduced_2024,he2025discoveringhighlyefficientlowweight}.
In particular, qLDPC codes (i.e., stabilizer code families with asymptotically $O(1)$ weight and degree) with good connectivity are theoretically understood to offer a compelling route toward fault tolerance with low overhead~\cite{10.5555/2685179.2685184, breuckmann_qldpc_2021, cohen_low-overhead_2022, iolius_almost-linear_2024,tamiya2024polylogtimeconstantspaceoverheadfaulttolerantquantum} and have garnered surging attention in QEC and fault tolerance.
Remarkably, the implementation of general qLDPC codes aligns well with current experimental advances  with  e.g.,~the reconfigurable atom array platform~\cite{Bluvstein_2023}, offering compelling prospects for highly efficient practical fault tolerance based on them.

The study of qLDPC codes has largely focused on the theoretical setting where the asymptotic parameter scalings in the infinite-code-length limit is of interest (see e.g.~Ref.~\cite{breuckmann_qldpc_2021} for an early review and Refs.~\cite{panteleev2022asymptoticallygoodquantumlocally,Leverrier_2022,dinur2022goodquantumldpccodes} for recent breakthroughs achieving qLDPC code families with asymptotically optimal rate and distance).
However, from a practical perspective, more relevant are explicit finite-size code parameters and constructions, the theory and optimization of which present markedly different problems with limited systematic understanding so far.
Crucially, extending the rich results on code parameter bounds~\cite{knill1997theory,gottesman1996hamming,gottesman1997stabilizer,ashikhmin1999upper,BravyiTerhal_2009,BPT,Nielsen_Chuang_2010,gottesman2016surviving},  the joint limits and achievability of weight and conventional code parameters remain little understood outside the asymptotic regime.

To understand finite-size stabilizer codes under weight constraints, we first address the natural, fundamental question: what is the minimum check weight required to realize an $[\![n,k,d]\!]$ code?
Formally, the weight of a stabilizer code is defined by the smallest possible maximum generator weight optimized over all generating sets of its stabilizer group $G$.
We denote the minimum weight for $[\![n,k,d]\!]$ stabilizer codes by $\wopt(n,k,d)$.
First, we prove that computing the stabilizer code weight is $\mathsf{NP}$-hard and thus computationally intractable, motivating the development of provable or efficiently computable bounds.
We establish a blocklength--weight tradeoff: for fixed $k\ge1$ and $d\ge2$, the optimal check weight is nonincreasing with blocklength and eventually reaches $3$ for $d=2$ or $4$ for $d\ge3$.
We then derive analytical lower bounds on check weight in terms of code rate and distance for arbitrary finite-length stabilizer codes, with stronger bounds for $d\ge2$, $d\ge3$, and $d\ge4$.
For CSS codes, we derive a stronger bound and give matching constructions at distance three.
At check weight at most $3$, we establish the sharp limits $d=2$ and $k/n\le1/4$ for codes with $k\ge1$ and $d\ge2$.
Furthermore, we develop a linear-programming (LP) method based on quantum weight enumerators, extending standard LP feasibility conditions with new constraints from bounded-weight generators, including growth laws, overlap, and parity constraints.
This yields an efficient algorithm to lower-bound $\wopt$ and gives exact optimal weights, with matching constructions, for all parameter triples with $n\le9$.
Finally, we show that our LP framework can be adapted to incorporate physical architectural constraints.
Applying a natural check-placement rule to the connectivity graph of the 127-qubit IBM Eagle processor~\cite{Kim2023Evidence,CarreraVazquez2024Combining}, our method certifies that check neighborhoods of radius $r\ge5$ are necessary to realize a $[\![127,100,6]\!]$ stabilizer code.

\section{Optimal weight for stabilizer codes}
Consider stabilizer quantum codes, which are uniquely defined by the +1-eigenspace of stabilizer groups given by $(-I)$-free abelian subgroups of the multi-qubit Pauli group.
As a standard notation, a stabilizer code has parameters $[\![n,k,d]\!]$ where $n,k$ are the numbers of physical and logical qubits respectively, and $d$ is the code distance.
Of central interest in this work are weight parameters, which we now formally define.
\begin{defn}\label{defn:W_and_W_opt}
For a stabilizer group $G$, define
\begin{equation}
\mathrm{W}(G)\coloneqq
\min_{\substack{S\subset G\\ \langle S\rangle=G}}
\max_{s\in S}\wt(s),
\end{equation}
(where $\wt(s)$ denotes the number of non-identity tensor factors in $s$), namely, the optimal maximum-generator-weight of $G$ among all possible choices of generating sets.
For a parameter triple $(n,k,d)$, define the optimal weight by
\begin{equation}
\mathrm{W}_{\!\mathrm{opt}}(n,k,d)\coloneqq
\min_{\substack{G\text{ defines an }[\![n,k,d]\!]\text{ code}}}
\mathrm{W}(G).
\end{equation}
If no $[\![n,k,d]\!]$ stabilizer code exists, we set
$\mathrm{W}_{\!\mathrm{opt}}(n,k,d)=\infty$.
\end{defn}

An $[\![n,k,d]\!]$ stabilizer code is said to have parameters $[\![n,k,d;w]\!]$ if $w=\mathrm{W}(G)$, where $G$ is its stabilizer group.
Trivially, $\mathrm{W}_{\!\mathrm{opt}}(n,k,d)=1$ when $d=1$ or $k=0$; henceforth, we focus on the nontrivial regime $k\ge 1$ and $d\ge 2$.

A fundamental question is the computational complexity of the weight parameter.
Computing $\w(G)$ resembles an integer program, suggesting  $\mathsf{NP}$-hardness.
Indeed, we prove (formal proof presented in Appendix~\ref{app:complexity}):
\begin{thm}\label{thm:W_NPhard}
The computation of $\mathrm{W}(G)$ is $\mathsf{NP}$-hard.
\end{thm}
The $\mathsf{NP}$-hardness of computing $\w(G)$ also suggests that determining $\wopt(n,k,d)$ is generally hard to compute, since it involves minimizing $\w(G)$ over all $[\![n,k,d]\!]$ stabilizer codes.
Consequently, to probe the value of $\wopt$, it is essential to develop both upper bounds (via explicit code constructions) and lower bounds (as we do in this work) for it.

Using the theoretical and LP bounds derived in the following sections, we obtain Algorithm~\ref{algo:Weight_LB_Table} (in Appendix~\ref{app:Pseudo_code}), which runs in time $\mathrm{poly}(N)$ and outputs lower bounds $\wlb(n,k,d)\le \wopt(n,k,d)$ for all parameter triples with $n\le N$.
Table~\ref{table:n_4_to_9} and Table~\ref{table:n_10_to_12} in Sec.~\ref{sec:finite_length_bounds_constructions} summarize our results for $\wopt(n,k,d)$ for $n\le 12$, where for $n\le 9$ we obtain the exact values.

To achieve some target $(k,d)$, one must choose a blocklength $n$ and a set of stabilizer checks.
As $n$ increases, it is often possible to realize the same $(k,d)$ with smaller check weight, reflecting a tradeoff between code efficiency and weight.
Typically, higher-weight checks incur larger measurement overhead (e.g., more ancillas and/or deeper circuits).
This raises a natural question: as $n$ grows, what is the minimum check weight required?
We prove the following theorem.
\begin{thm}\label{thm:weight-floor}
For any $k\ge1$, we have:
\begin{enumerate}[label=(\roman*)]
\item $\wopt(n',k,d)\le \wopt(n,k,d)$ for $n'\ge n$.
\item $\min_{n\ge1}\wopt(n,k,2)=3$.
\item For $d\ge3$, $\min_{n\ge1}\wopt(n,k,d)=4$.
\end{enumerate}
\end{thm}
\begin{proof}
(i)
For any $[\![n,k,d;w]\!]$ code we can add $n'-n$ qubits with 1-qubit $Z$-checks on them to get a  $[\![n',k,d;w]\!]$ code.

(ii)
By Theorem~\ref{thm:w_LBby_nk}, for any code with $k\ge1,d\geq 2$, the maximum weight $\geq 3$. There exists a $[\![4,1,2;3]\!]$ code defined by stabilizer group $\langle XXXI,IYYY,ZIZZ \rangle$. For any $k$, the tensor product of $k$ copies of this code would be a construction.

(iii)
For $d\ge3$, the rotated surface code~\cite{eczoo_rotated_surface} on $d\times d$ lattice has parameters $[\![d^2,1,d;4]\!]$.
The tensor product of $k$ copies of this code has parameters $[\![kd^2,k,d;4]\!]$. By Theorem~\ref{thm:wle3_main}, weight-3 codes could not have distance $\ge 3$.
\end{proof}

To illustrate how the optimal weight decreases and finally reaches the limit of $3$ or $4$ guaranteed by the above theorem, consider the following two choices of $(k,d)$.
For $(k,d)=(5,2)$, we have
\begin{equation}
\operatorname{W}_{\!\operatorname{opt}}(n,5,2)=
\begin{cases}
\infty,&n\le7,\\
6,&n=8,\\
5,&n=9,\\
4,&10\le n\le19,\\
3,&n\ge20,
\end{cases}
\end{equation}
whereas for $(k,d)=(3,3)$,
\begin{equation}
\operatorname{W}_{\!\operatorname{opt}}(n,3,3)=
\begin{cases}
\infty,&n\le7,\\
6,&n=8,\\
5,&n=9,\\
4,&n\ge15.
\end{cases}
\end{equation}
\subsection*{Average generator weight}
While we primarily focus on the maximum generator weight in this paper, the average generator weight is also practically important.
For a nontrivial stabilizer group $G$, we minimize the average weight over independent generating sets and define
\begin{align}
\wavg(G)&\coloneqq\min_{\substack{S\subset G,\ \langle S\rangle=G,\\ |S|=\log_2|G|}}\frac{1}{|S|}\sum_{s\in S}\wt(s),\\
\woptavg(n,k,d)&\coloneqq\min_{G\text{ has parameters }[\![n,k,d]\!]}\wavg(G).
\end{align}
Remarkably, $\w$ and $\wavg$ can be optimized simultaneously, through a ``weight-optimal" generating set:
\begin{defn}\label{def:weight-optimal_generating_set}
Let $G$ be a stabilizer group with $\abs{G} = 2^r$.
Let $S=\{g_1,\cdots,g_r\}\subset G$ be a generating set ordered so that
$\wt(g_1)\le \cdots \le \wt(g_r)$.
We say that $S$ is \emph{weight-optimal}\footnote{For every stabilizer group, there exists a weight-optimal generating set; see Appendix~\ref{app:Existence_weight_optimal}.
Any two weight-optimal generating sets $\{g_1^{(1)},\cdots,g_{r}^{(1)}\}$ and $\{g_1^{(2)},\cdots,g_{r}^{(2)}\}$ of $G$ satisfy $\wt(g_i^{(1)})=\wt(g_i^{(2)})$, for all $i\in[r]$.
By Theorem~\ref{thm:W_NPhard}, finding a weight-optimal generating set is also $\mathsf{NP}$-hard, since it directly yields $\w(G)$.} if, for every generating set
$\{h_1,\cdots,h_r\}\subset G$ of $G$ ordered so that $\wt(h_1)\le \cdots \le \wt(h_r)$, one has
\begin{equation}
\wt(g_i)\le \wt(h_i),\quad i=1,\cdots,r.
\end{equation}
\end{defn}

\begin{prop}\label{prop:optimal_generators}
Let $S$ be a weight-optimal generating set of a stabilizer group $G$, then the maximum and average weight of $S$ simultaneously attain the optimal value. That is, $\max_{s\in S}\wt(s)=\w(G)$ and $\frac{1}{|S|}\sum_{s\in S}\wt(s)=\wavg(G)$.
\end{prop}

However, for a target parameter $[\![n,k,d]\!]$, optimizing the maximum weight $\wopt$ does not necessarily optimize the average weight $\woptavg$ (and vice versa), so these two objectives can be in tension.
We illustrate this mismatch for $[\![4,1,2]\!]$ codes.
Consider two stabilizer groups
\begin{align}
G_1=&\;\langle XXXI,IYYY,ZIZZ\rangle,\\
G_2=&\;\langle ZZII,IIZZ,XXXX\rangle,
\end{align}
they form two $[\![4,1,2]\!]$ codes.
$\wopt(4,1,2)=3$ is achieved on $G_1$, $\woptavg(4,1,2)=8/3$ is achieved on $G_2$. Yet we have
\begin{align}
\w(G_1)=3&<4=\w(G_2),\\
\wavg(G_1)=3&>8/3=\wavg(G_2),
\end{align}
and the optimal value for $\wopt$ and $\woptavg$ cannot be obtained on the same stabilizer group.
\section{Fundamental bounds on check weight}
We derive lower bounds on check weight in terms of code rate and distance.
These bounds are also helpful in reducing the computational cost of the LP bounds below, as they allow us to exclude theoretically forbidden parameter sets $[\![n,k,d;w]\!]$ without solving the LP.

For CSS codes, define $\wopt^{\mathrm{CSS}}(n,k,d)$ as the minimum of $\w(G)$ over all CSS stabilizer groups $G$ defining an $[\![n,k,d]\!]$ code. In Appendices~\ref{app:proof_nk_bound}--\ref{app:analytic_rate_bounds}, we prove:

\begin{thm}[Lower bounds on check weight]\label{thm:w_LBby_nk}\label{thm:rate_weight_main}\label{thm:css_rate_weight_main}
For any integers $n>k\ge1$, the optimal check weight satisfies
\begin{align}
\wopt(n,k,d) &\ge \left\lceil\frac{2n}{n-k}\right\rceil,\quad \text{for } d\ge2,\label{eq:d2_weight_floor_main}\\
\wopt(n,k,d) &\ge \max\!\left\{4,\left\lceil\frac{11n+k+1}{4(n-k)}\right\rceil\right\},\quad \text{for } d\ge3,\label{eq:d3_weight_floor_main}\\
\wopt(n,k,d) &\ge \max\!\left\{4,\left\lceil\frac{3n}{n-k}\right\rceil\right\},\quad \text{for } d\ge4.\label{eq:d4_weight_floor_main}
\end{align}
For CSS codes with $d\ge3$,
\begin{equation}\label{eq:css_weight_floor_main}
\wopt^{\mathrm{CSS}}(n,k,d)\ge\left\lceil\frac{3n+k}{n-k}\right\rceil.
\end{equation}
\end{thm}

\begin{figure}[tbp]
\centering
\includegraphics[width=0.9\columnwidth]{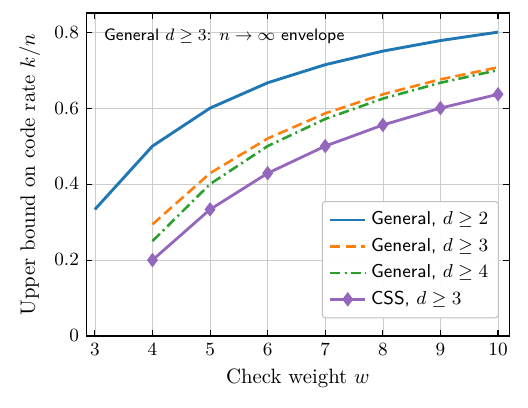}
\caption{Rate bounds from Theorem~\ref{thm:w_LBby_nk}.
Curves show upper bounds on $k/n$ as functions of check weight $w$.
The general $d\ge3$ curve omits the $O(1/n)$ finite-length correction.
Filled diamonds mark complete-graph CSS codes attaining the CSS bound.}
\label{fig:rate_weight_bounds}
\end{figure}

The corresponding rate bounds are illustrated in Fig.~\ref{fig:rate_weight_bounds}.

To see \eqref{eq:d2_weight_floor_main}, note that when $d\ge2$, every effective qubit used to encode logical information must be covered by at least two stabilizer generators with distinct Paulis on it, requiring $w(n-k)\ge2n$.
As an immediate example, for $m\ge2$, a $[\![2m,2m-2,2]\!]$ code must have maximum generator weight $\ge 2m$. Consequently, the code with stabilizer group $\langle X^{\otimes 2m},\, Z^{\otimes 2m}\rangle$ is optimal.
The bound \eqref{eq:d3_weight_floor_main} follows by refining the incidence count using commutation and the fact that, for $d\ge3$, any weight-$2$ Pauli commuting with all checks must be a stabilizer.
Counting these stabilizers constrains the number of qubits covered by only two checks.
For $d\ge4$, weight-$3$ commuting Paulis must also be stabilizers, strengthening the count to give \eqref{eq:d4_weight_floor_main}.
For CSS codes, counting the $X$- and $Z$-check incidences separately yields the stronger bound \eqref{eq:css_weight_floor_main}.
The CSS bound \eqref{eq:css_weight_floor_main} is attained for every integer $w\ge4$ by $[\![w(w+1)/2,w(w-3)/2,3;w]\!]$ codes, constructed from self-dual cellular embeddings of complete graphs~\cite{naghipour2017classes,archdeacon1994selfdual}.
Appendix~\ref{app:analytic_rate_bounds} gives the construction and verifies its parameters and optimality.

For $k\ge1$ and $d\ge2$, \eqref{eq:d2_weight_floor_main} implies that the maximum generator weight is at least $3$, that is, any weight $<3$ code is trivial.
We now focus on codes attaining this minimum.
Commutation restricts the overlaps of weight-$3$ checks, ruling out distance $d\ge3$ for codes with $k\ge1$.
To bound the rate, we remove fixed one-qubit factors and count the incidences between independent checks and the remaining qubits.
Each remaining qubit is covered by at least two checks, and every weight-$3$ check contains a qubit covered by at least three checks.
These constraints give $n\ge4k$.
The complete proof of the following theorem is given in Appendix~\ref{app:wle3}.

\begin{thm}\label{thm:wle3_main}
Suppose $k\ge1$, $d\ge2$.
If there exists an $[\![n,k,d;3]\!]$ code, then $d=2$ and $k/n\le1/4$.
\end{thm}

Consequently, if $d=2$ and $k/n>1/4$, then $\wopt(n,k,d)\ge4$.
For distance $2$ with $k/n\le1/4$, the $k$-fold tensor product of the $[\![4,1,2;3]\!]$ block, padded by $n-4k$ trivial qubits, gives an $[\![n,k,2;3]\!]$ code and hence attains $\wopt(n,k,2)=3$.
The $[\![5,1,3]\!]$ perfect code and the 2D toric code provide weight-$4$ examples saturating the weight-$4$ floor for $d\ge3$.

\section{Distance limitations at low weight}
Larger check weight generally allows larger distance.
The very low-weight regime is rigid: $w\le 2$ forces $d=1$, and $w=3$ gives $d\le 2$ by Theorem~\ref{thm:wle3_main}.
At the positive end, good qLDPC codes are known with check weight $6$, via recent layer-code weight reduction applied to existing good CSS qLDPC families~\cite{panteleev_quantum_2022,Leverrier_2022,yuan2026quantum}.
This leaves a narrow gap at $w=4,5$.
For weight-$4$ and weight-$5$ codes, Theorem~\ref{thm:rate_weight_main} gives asymptotic rate ceilings of $5/17$ and $3/7$ for $d\ge3$, tightening to $1/4$ and $2/5$ for $d\ge4$, respectively.
The case $w=5$ is subtle: coning and related weight-reduction approaches~\cite{hastings_weight_2016,hastings2021quantum,hsieh2025simplifiedquantumweightreduction} suggest a possible route to this regime.
For weight-$4$ codes, growing distance is possible, but we conjecture that every $[\![n,k,d;4]\!]$ code satisfies the square-root barrier $d=O(\sqrt n)$.
This conjectured barrier is tight for standard weight-$4$ product-type examples, including hypergraph-product codes and the toric code, which achieve $d=\Theta(\sqrt n)$~\cite{Zemor2014,Leverrier_2015,Zeng_2019,Zeng_2020}.
All currently established constructions that surpass the square-root distance barrier rely on higher-dimensional expander or product structure with check weight exceeding $4$~\cite{10.1145/3406325.3451005,panteleev_quantum_2022,breuckmann_balanced_2021,panteleev2022asymptoticallygoodquantumlocally,Leverrier_2022,dinur2022goodquantumldpccodes,Dinur2024sheaf}, providing evidence for the conjecture.

\section{Bounds from linear programming}
\emph{Weight enumerators} provide a convenient way to examine the weight structure of  codes~\cite{gottesman2016surviving}.
Here we demonstrate how to apply LP methods based on enumerators to the weight-constrained situation so as to systematically produce refined code parameter bounds especially in the finite-size regime.

Given a quantum code defined by projector $\Pi$ onto a rank-$K$ space ($K=2^k$ for stabilizer codes), define the weight enumerators $A_0,\cdots,A_n$ by (up to normalization) summing the squared overlaps $\abs{\Tr(P\Pi)}^2$ over all weight-$j$ Pauli operators $P\in\{I,X,Y,Z\}^{\otimes n}$.
Intuitively, $A_j$ measures how strongly the code space ``correlates" with weight-$j$ Paulis, and for stabilizer codes it reduces to a counting function~\cite{gottesman2016surviving}:
\begin{equation}\label{eq:stabilizer_code_Aj}
A_j=\abs{\{g\in G:\wt(g)=j\}}.
\end{equation}
Two related distributions, dual enumerators $B_j$ and shadow enumerators $Sh_j$, encode further positivity constraints and dual information.

These quantities are  related: the quantum MacWilliams identity relates $\vec B$ and $\vec A$ through a \emph{Krawtchouk matrix} (a fixed $(n+1)\times(n+1)$ matrix $M$)~\cite{Shor1997MacWilliams}, and similarly $\vec{Sh}$ can be obtained from $\vec A$ via a \emph{signed Krawtchouk matrix} $\widetilde M$~\cite{Rains1998Shadow,Rains1999shadow}.
Importantly, the distance $d$ is characterized by the condition $A_j=B_j$ for all $j<d$~\cite{gottesman2016surviving}.
Consequently, the existence of such a quantum code indicates the feasibility of an efficiently solvable linear program in the variables $A_0,\cdots,A_n$ with constraints: i) $A_0=B_0=1$, ii) $B_j\ge A_j\ge0$, iii) $A_j=B_j$ for $j<d$, and iv) $Sh_j\ge0$.
Infeasibility of this LP certifies that no such code exists.
See Appendix~\ref{app:enumerator_and_LP} for more detailed information.

We now incorporate the weight parameter.
If an $[\![n,k,d;w]\!]$ stabilizer code exists, then one can derive additional linear constraints by linking $w$ to the weights of stabilizer-group elements, as captured by  $A_j$.
A key observation is that low-weight generators force many stabilizer elements to have small weight.
Since there are $n-k$ independent generators of weight at most $w$, \eqref{eq:stabilizer_code_Aj} implies $\sum_{j=1}^{w}A_j\ge n-k$; Moreover, the product of any two generators has weight at most $2w$, hence $\sum_{i=1}^{2w}A_i\ge (n-k)+\binom{n-k}{2}$. More generally, for each $C=1,\cdots,\lfloor n/w\rfloor$ we obtain
\begin{equation}
\sum_{i=1}^{Cw}A_i\ge \sum_{c=1}^{C}\binom{n-k}{c}.
\end{equation}
These cumulative lower bounds provide a coarse ``growth law" for the weights of stabilizer-group elements.

In Appendix~\ref{app:linear_constraints_weight} we sharpen these constraints in several directions. First, we track how many generators attain the maximal weight $w$, which yields additional lower/upper bounds on $A_w$ and $\sum_{j<w}A_j$, as well as tighter cumulative lower bounds.
Second, for certain parameter candidates $[\![n,k,d;w]\!]$, the code cannot decompose as a tensor product of smaller codes (which is computationally efficient to identify); hence, $A_1=0$, and its generators need to have more overlap to be ``connected".
Finally, we incorporate parity constraints valid for stabilizer codes: in any stabilizer group, either all elements have even weight, or even/odd weights are balanced.
These refinements make the resulting LP bounds substantially tighter in practice.
We organize these constraints into the pseudocode in Algorithm~\ref{algo:Weight_LB_Table} in Appendix~\ref{app:Pseudo_code}. Using this procedure, we can generate in $\mathrm{poly}(N)$ time lower bounds $\wlb(n,k,d)$ for $\wopt(n,k,d)$ for all $n\le N$.

\begin{thm}[No-go criteria from LP family (informal)]\label{thm:WLB_LP_certificate_main}
If an $[\![n,k,d;w]\!]$ stabilizer code exists, then its weight enumerators $(A_0,\cdots,A_n)$ must satisfy the standard LP constraints for $[\![n,k,d]\!]$ codes, together with the additional weight-induced constraints summarized in Appendix~\ref{app:linear_constraints_weight}, under ``Summary'' (for some admissible choice of auxiliary discrete parameters).
\end{thm}
Consequently, if the weight-induced constraints cannot be satisfied for all admissible choices of parameters, then no $[\![n,k,d;w]\!]$ stabilizer code can exist. See Appendix~\ref{app:linear_constraints_weight} for the formal version of Theorem \ref{thm:WLB_LP_certificate_main}.

This method leads to lower bounds for $\wopt$ in Table~\ref{table:n_4_to_9} ($n\le9$) and Table~\ref{table:n_10_to_12} ($10\le n\le 12$). In Table~\ref{table:n_4_to_9}, all lower bounds match the corresponding upper bounds and are thus tight.
In Table~\ref{table:n_10_to_12}, an entry of the form $a$--$b$ means that our lower and upper bounds currently satisfy $a\le \wopt(n,k,d)\le b$.
We solved the LPs in Mathematica~\cite{Mathematica} using exact (rational) arithmetic, thereby avoiding floating-point round-off errors.

\section{Practical code architectures}
While we have so far focused on general bounds, actual experimental setups or architectures most likely give rise to more specific connectivity constraints, such as geometric locality, on the check configuration. This clearly motivates the study of situations subject to such more refined conditions.

\begin{figure}[tbp]
\centering
\includegraphics[width=\columnwidth]{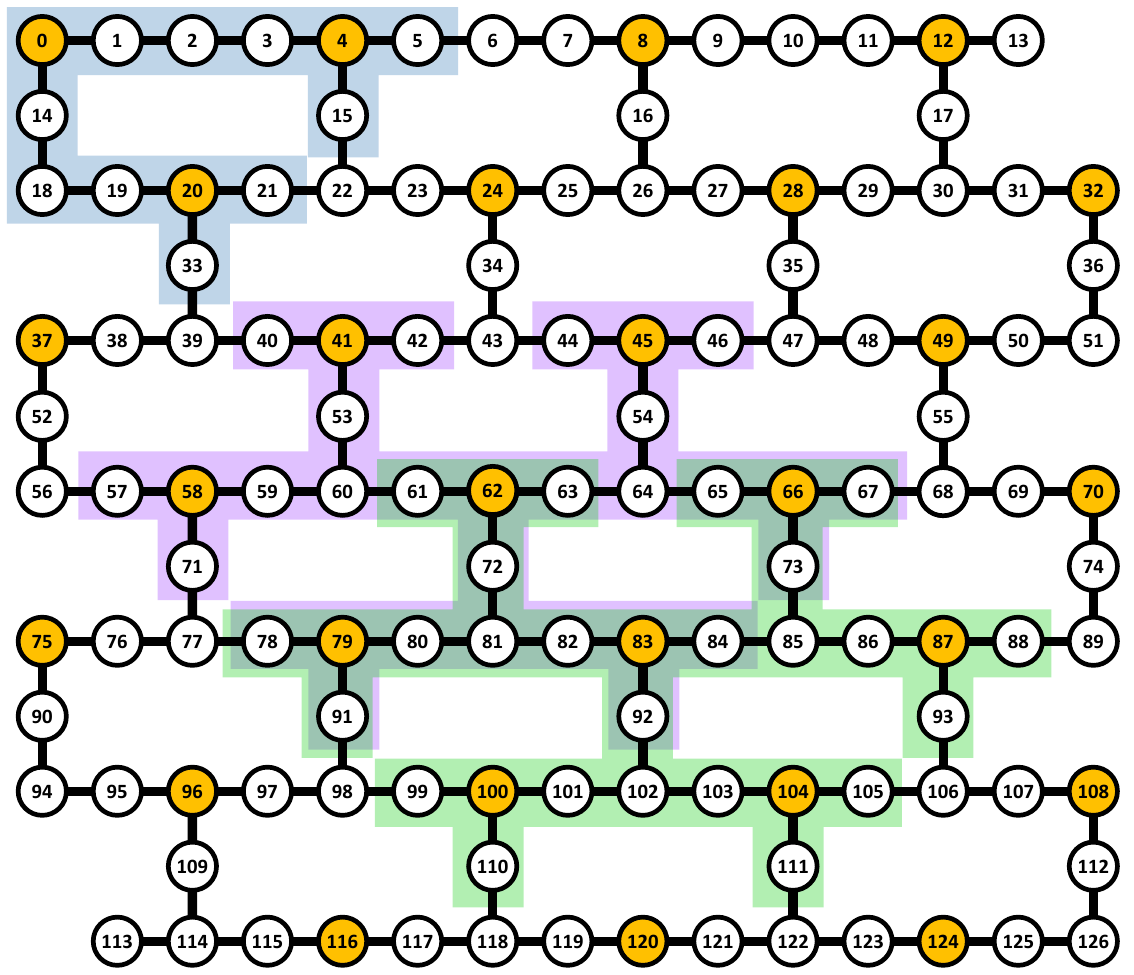}
\caption{Qubit layout of the IBM \texttt{Eagle} chip. With checks centered at the 27 qubits colored in orange, our LP lower bound implies that radius $r\ge5$ is required to achieve a $[\![127,100,6]\!]$ code. The three shaded areas show the radius-5 neighborhoods of qubits 0, 62, and 83.}
\label{fig:IBM127}
\end{figure}

As a representative practical example, consider the IBM 127-qubit ``\texttt{Eagle}" chip~\cite{Kim2023Evidence,CarreraVazquez2024Combining}, whose qubit connectivity graph is shown in Fig.~\ref{fig:IBM127}. We ``evenly" place Pauli check operators according to the hardware layout as follows: on each of the 27 qubits colored orange in Fig.~\ref{fig:IBM127}, we assign a check supported within the radius-$r$ neighborhood of that qubit in the connectivity graph.

To illustrate our architecture-dependent analysis, consider the
natural case $(n,k)=(127,100)$, for which stabilizer codes with distance $d=6$ are known to exist~\cite{grasslCodetables}.
Now the question is the following: under the above check-placement rule, what radius $r$ is necessary to realize a $[\![127,100,6]\!]$ stabilizer code on the \texttt{Eagle} chip?

We can use LP to lower bound $r$.
Without exploiting the geometric information of check locations, imposing the coarse constraint $\sum_{i=1}^{Cw}A_i\ge \sum_{c=1}^{C}\binom{n-k}{c}$ yields a lower bound of $13$ on the maximum stabilizer weight.
Since the central checks reach weight $13$ at radius $r=3$, this baseline bound gives $r\ge3$.

Now taking the hardware layout into account, we can strengthen the LP constraints and obtain the improved bound
\begin{equation}
r \ge 5.
\end{equation}
The key insight is that the prescribed geometry allows us to derive much sharper upper bounds on the weights of products of checks by accounting for overlaps of their supports. For instance, radius-$5$ neighborhoods of the qubits 62 (purple) and 83 (green) overlap substantially, so the product of the corresponding checks satisfies $\wt(g_{62}g_{83})\le\mathrm{ub}(g_{62}g_{83})\coloneqq 2\times 31-\#\mathrm{overlap}=45$, which is significantly better than the structure-agnostic bound $62$.
Generally, by enumerating the support-union upper bounds $\mathrm{ub}(g)$ for all elements $g\in G$ generated by these 27 checks (taking overlaps into account), we obtain, for every $M=0,1,\cdots,n$, the cumulative constraints
\begin{equation}
\sum\nolimits_{i\le M}A_i\ge\#\{g\in G:\mathrm{ub}(g)\le M\}.
\end{equation}
After adding these constraints, the resulting LP remains infeasible for $r\le 4$, and becomes feasible at $r=5$.

\section{Finite-length low-weight constructions and bounds}\label{sec:finite_length_bounds_constructions}

\begin{table*}
\centering
\begin{tabular}{c|c}
\hline
$\mathbf{n=4}$ & $d=2$ \\
\hline
$k=1$ & 3 \\
\hline
$k=2$ & 4 \\
\hline
\end{tabular}
\;
\begin{tabular}{c|c|c}
\hline
$\mathbf{n=5}$ & $d=2$ & $d=3$ \\
\hline
$k=1$ & 3 & 4 \\
\hline
$k=2$ & 4 & $\infty$ \\
\hline
\end{tabular}
\;
\begin{tabular}{c|c|c}
\hline
$\mathbf{n=6}$  & $d=2$ & $d=3$ \\
\hline
$k=1$ & 3 & 4 \\
\hline
$k=2$ & 4 & $\infty$ \\
\hline
$k=3$ & 4 & $\infty$ \\
\hline
$k=4$ & 6 & $\infty$ \\
\hline
\end{tabular}
\;
\begin{tabular}{c|c|c}
\hline
$\mathbf{n=7}$ & $d=2$ & $d=3$ \\
\hline
$k=1$ & 3 & 4 \\
\hline
$k=2$ & 4 & $\infty$ \\
\hline
$k=3$ & 4 & $\infty$ \\
\hline
$k=4$ & 6 & $\infty$ \\
\hline
\end{tabular}
\;
\begin{tabular}{c|c|c}
\hline
$\mathbf{n=8}$ & $d=2$ & $d=3$ \\
\hline
$k=1$ & 3 & 4 \\
\hline
$k=2$ & 3 & 4 \\
\hline
$k=3$ & 4 & 6 \\
\hline
$k=4$ & 4 & $\infty$ \\
\hline
$k=5$ & 6 & $\infty$ \\
\hline
$k=6$ & 8 & $\infty$ \\
\hline
\end{tabular}
\;
\begin{tabular}{c|c|c}
\hline
$\mathbf{n=9}$ & $d=2$ & $d=3$ \\
\hline
$k=1$ & 3 & 4 \\
\hline
$k=2$ & 3 & 4 \\
\hline
$k=3$ & 4 & 5 \\
\hline
$k=4$ & 4 & $\infty$ \\
\hline
$k=5$ & 5 & $\infty$ \\
\hline
$k=6$ & 7 & $\infty$ \\
\hline
\end{tabular}
\caption{The value of $\wopt(n,k,d)$ for all codes with $n\le 9$. Lower bounds are obtained via Algorithm~\ref{algo:Weight_LB_Table} in Appendix~\ref{app:Pseudo_code}. Constructions achieving these bounds (i.e., weight-optimal codes) are provided in Appendix~\ref{app:Low-weight_code_constructions}. The cases $10\le n\le 12$ are listed separately in Table~\ref{table:n_10_to_12}.}
\label{table:n_4_to_9}
\end{table*}

\begin{table*}
\centering
\begin{tabular}{c|c|c|c}
\hline
$\mathbf{n=10}$ & $d=2$ & $d=3$ & $d=4$ \\
\hline
$k=1$ & 3 & 4 & 4 - 6 \\
\hline
$k=2$ & 3 & 4 & 5 - 6 \\
\hline
$k=3$ & 4 & 5 & $\infty$ \\
\hline
$k=4$ & 4 & 6 - 7 & $\infty$ \\
\hline
$k=5$ & 4 & $\infty$ & $\infty$ \\
\hline
$k=6$ & 6 & $\infty$ & $\infty$ \\
\hline
$k=7$ & 7 & $\infty$ & $\infty$ \\
\hline
$k=8$ & 10 & $\infty$ & $\infty$ \\
\hline
\end{tabular}
\;
\begin{tabular}{c|c|c|c|c}
\hline
$\mathbf{n=11}$ & $d=2$ & $d=3$ & $d=4$ & $d=5$\\
\hline
$k=1$ & 3 & 4 & 4 - 6 & 6 \\
\hline
$k=2$ & 3 & 4 & 4 - 6 & $\infty$ \\
\hline
$k=3$ & 4 & 4 - 5 & $\infty$ & $\infty$ \\
\hline
$k=4$ & 4 & 5 - 7 & $\infty$ & $\infty$  \\
\hline
$k=5$ & 4 & 6 - 8 & $\infty$ & $\infty$  \\
\hline
$k=6$ & 5 & $\infty$ & $\infty$ & $\infty$  \\
\hline
$k=7$ & 6 & $\infty$ & $\infty$ & $\infty$  \\
\hline
$k=8$ & 8 & $\infty$ & $\infty$ & $\infty$  \\
\hline
\end{tabular}
\;
\begin{tabular}{c|c|c|c|c}
\hline
$\mathbf{n=12}$ & $d=2$ & $d=3$ & $d=4$ & $d=5$\\
\hline
$k=1$ & 3 & 4 & 4 - 6 & 6 \\
\hline
$k=2$ & 3 & 4 & 4 - 6 & $\infty$ \\
\hline
$k=3$ & 3 & 4 - 5 & 6 - 8 & $\infty$ \\
\hline
$k=4$ & 4 & 5 - 7 & 7 - 8 & $\infty$  \\
\hline
$k=5$ & 4 & 6 - 8 & $\infty$ & $\infty$  \\
\hline
$k=6$ & 4 & 8 - 10 & $\infty$ & $\infty$  \\
\hline
$k=7$ & 6 & $\infty$ & $\infty$ & $\infty$  \\
\hline
$k=8$ & 6 & $\infty$ & $\infty$ & $\infty$  \\
\hline
$k=9$ & 8 & $\infty$ & $\infty$ & $\infty$  \\
\hline
$k=10$ & 12 & $\infty$ & $\infty$ & $\infty$  \\
\hline
\end{tabular}
\caption{The values of $\wopt(n,k,d)$ for $n=10,11,12$.
Theorem~\ref{thm:rate_weight_main} yields the lower bounds $\wopt(10,3,3)\ge5$ and $\wopt(12,4,3)\ge5$. The lower bound $\wopt(12,7,2)\ge 6$ follows from the dedicated argument in Appendix~\ref{app:specific_parameter_bound}, while all remaining lower bounds on $\wopt$ are obtained via Algorithm~\ref{algo:Weight_LB_Table} in Appendix~\ref{app:Pseudo_code}.}
\label{table:n_10_to_12}
\end{table*}

\begin{figure*}
    \centering
    \includegraphics[width=0.72\textwidth]{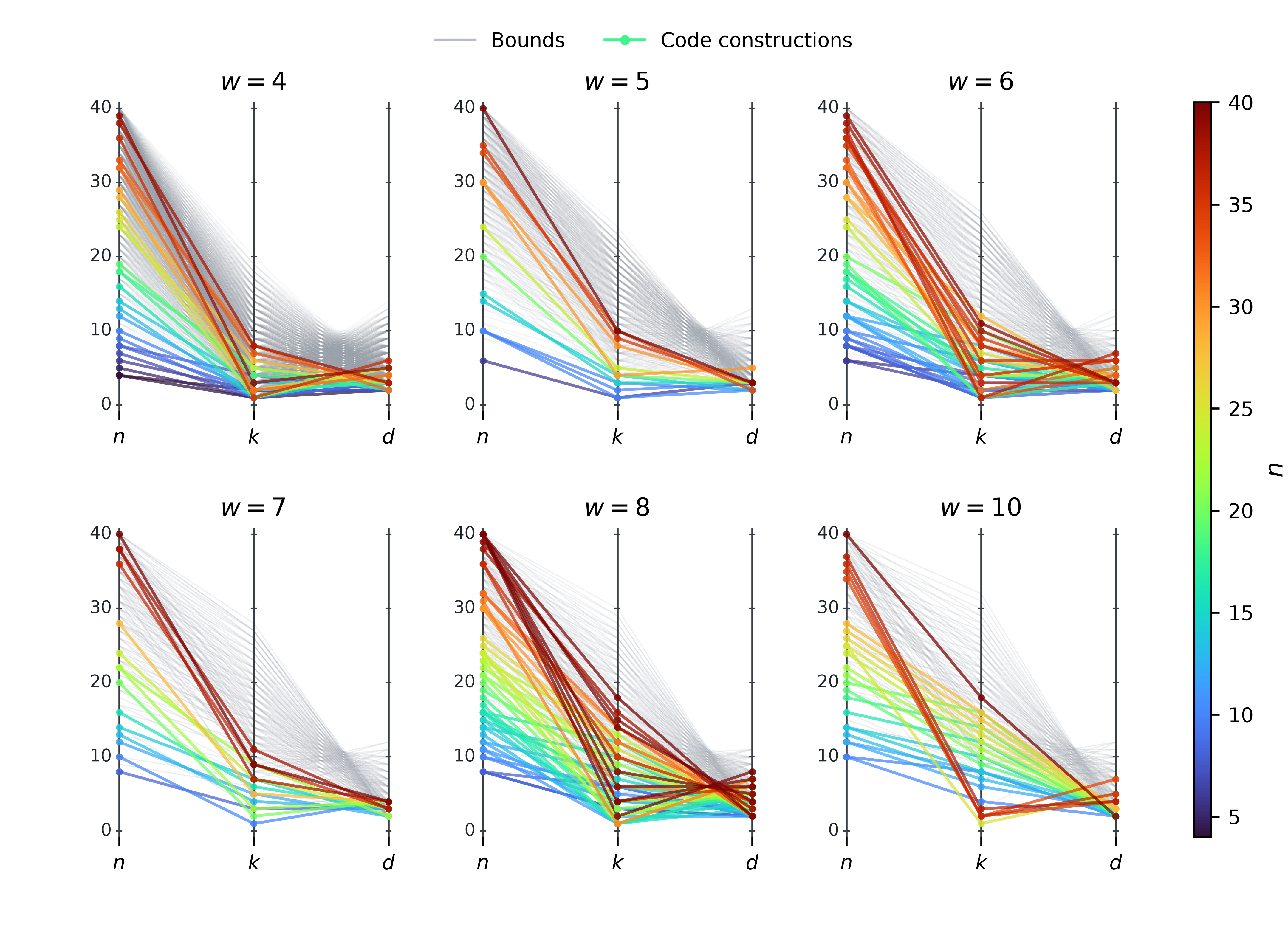}
    \caption{Parallel-coordinate comparison of our lower bounds with explicit finite-length stabilizer code constructions both from the literature and discovered by our reinforcement learning-based code design algorithms, at different weights $w$. Each polyline represents an $(n,k,d)$ triple. Gray lines illustrate lower-bound data, while colored lines represent explicit constructions, with color indicating the block length $n$.}
    \label{fig:parallelcoords}
\end{figure*}

To complement our lower bounds on $\mathrm{W}_{\!\mathrm{opt}}(n,k,d)$, we collect representative finite-length stabilizer codes with low check weight both found in the literature~\cite{grassl1997erasure,laflamme1996perfect,shaw2008six,jones2013multilevel,self2022iceberg,steane1996simple,gottesman1996hamming,gottesman1997stabilizer,calderbank1998gf4,calderbank1996orthogonal,grasslCodetables,kitaev2003fault,bombin2007optimal,bonilla2021xzzx,landahl2020rhombic,shor1995qec,srivastava2022xyz2,maheshwari2024eight,koukoulekidis2024small,paetznick2024carbon,liang2025twisted,liang2025selfdual,andersen2025small,lai2025constant,bombin2006topological,landahl2011color,bravyi2005universal,scott2005failure,khesin2024graph,hastings2025cyclic,koh2026phantom,bacon2006operator,conrad2018stellated,voss2024multivariate,vasmer2022morphing,wang2026coprime,berthusen2024fourD,jacob2025tricycle,liu2026stacked,khesin2026mirror,radebold2025tanner,tiew2026copycup,eczSmallCodes} as well as improved new examples obtained using our reinforcement learning-based code discovery scheme~\cite{he2025discoveringhighlyefficientlowweight}.
Together, the construction data and lower bounds bracket $\mathrm{W}_{\!\mathrm{opt}}(n,k,d)$.

We collectively illustrate the parameters of these state-of-the-art explicit constructions and our lower-bound data for different weights in Fig.~\ref{fig:parallelcoords}. At comparable block lengths, larger $k$ is generally associated with smaller $d$, revealing a clear rate--distance tradeoff at fixed check weight. Within the finite-size data, increasing $w$ expands the accessible parameter region primarily toward larger $k$, whereas the range of $d$ grows more modestly, showing that the check weight affects the rate more than the distance. The low-weight cases $w=4$ and $w=5$ consequently show the narrowest construction coverage.
Also, the discrepancy between the colored constructions and the gray lower-bound data is most pronounced for simultaneously large $k$ and $d$, highlighting this regime as the principal target for improved constructions.

A complementary rate--relative-distance representation of the finite-length lower-bound data is provided in Appendix~\ref{app:rate_distance}.

\section{Discussion}

Motivated by pressing theoretical questions and experimental needs in QEC and fault tolerance, we formulated and explored the theory of stabilizer codes under explicit check weight constraints. As our main contributions, we developed a variety of fundamental results concerning the weight parameter of stabilizer codes, addressing its computational hardness and integrating it into the theory of code parameter interplay to establish a systematic understanding of both the efficiency and limitations of low-weight codes.

Several problems are interesting directions for further study. Some entries in the parameter-bound tables are still unmatched, calling for stronger lower-bound methods, such as SDP relaxations~\cite{munné2025sdpboundsquantumcodes}, as well as improved explicit constructions. Another interesting direction is to investigate the average generator weight, particularly the observed tension between optimizing maximum and average weight. Finally, the framework can be extended to additional practical constraints, such as hardware connectivity, qubit degree, locality, and syndrome extraction, and specialized to structured code families such as HGP and other product-code constructions.

A broader long-term goal is to incorporate even more code properties---including logical operations, decoding features, check soundness and energy barriers---into the picture. Furthermore, extending the questions and methods here to more general dynamical and spacetime code settings is expected to be valuable. These efforts would offer more comprehensive insights and guidance for both the theoretical and experimental developments in quantum coding and fault tolerance.

\section*{Acknowledgements}
We thank Kun Fang, Yuguo Shao, and Bochen Tan for valuable discussion.
F.W. is supported by the Shuimu Tsinghua Scholar Program.
Z.H., Z.L., Z.-W.L. are supported in part by NSFC under Grant No.~12475023, Dushi Program, and startup funding from YMSC.

\emph{Note added.}
While finalizing this work, we noticed an excellent concurrent paper by Wang et al.~\cite{wang2026checkweightconstrainedquantumcodesbounds} that studies closely related questions.
The two works were carried out independently and are consistent where they overlap.
Key differences include the following.
Wang et al.~\cite{wang2026checkweightconstrainedquantumcodesbounds} considered the CSS and subsystem code settings more carefully, obtaining bounds for some special weight-$4$ CSS codes as well as weight-$2$ subsystem codes.
They also worked out specific instances of quantum Tanner codes, which provide additional insight into the finite-size behaviors of asymptotic codes.
We worked in the general stabilizer-code setting and further presented computational-complexity results, code rate condition for weight $3$, more refined LP constraints (which leads to various better bounds and, in particular, tight bounds for small $n$), and architecture-specific considerations.


%

\appendix
\onecolumngrid

\section{Notations and auxiliary lemmas}\label{app:notations}

Let $\mc{P}_n$ be the $n$-qubit Pauli group and let $\hat{\mc{P}}_n=\mc{P}_n/\langle iI\rangle$, whose elements are the $4^n$ Pauli strings with phase $+1$.
For any $P\in\mc{P}_n$, let $\hat{P}$ be the corresponding element in $\hat{\mc{P}}_n$. For a subset $A\subset\mc{P}_n$, denote $\hat{A}=\{\hat{a}: a\in A\}$. Let $N(A)=\{P\in\mc{P}_n:  PQ=QP,\forall Q\in A\}$ be the normalizer of $A$.
A stabilizer code with stabilizer group $G$ has distance $\min\{\wt(E):  E\in \hat{N}(G) \setminus \hat{G}\}$.
By definition, any $[\![n,0]\!]$ stabilizer code (i.e., a stabilizer state) satisfies $\hat{N}(G)=\hat{G}$, thus it has distance $\min\emptyset=\infty$.

For a Pauli string $P$ acting on $n$ qubits and a subset of qubits $S\subset[n]\coloneqq \{1,\cdots,n\}$, we write $P|_S$ as the Pauli string of $P$ restricting on $S$, and ignore the phase, e.g., $(-X\otimes Y\otimes Z)|_{\{1,3\}}=X\otimes Z$, $(-X\otimes Y\otimes Z)|_2=Y$. For two Pauli strings $P,Q$, denote $\{P,Q\}\coloneqq PQ+QP$.
We sometimes omit the $\otimes$, e.g. write $X\otimes Y\otimes Z$ as $XYZ$.

\begin{lem}\label{lemma:tensor_product_parameters}
Let $G_i\subset \mc{P}_{n_i}$ be stabilizer groups that define $[\![n_i,k_i,d_i;w_i]\!]$ stabilizer codes for $i=1,2$. Then the tensor product
$G\coloneqq G_1\otimes G_2=\{g_1\otimes g_2: \ g_i\in G_i\}\subset \mathcal P_{n_1+n_2}$
defines a stabilizer code with parameters
\begin{equation}
[\![n_1{+}n_2, k_1{+}k_2,\min\{d_1,d_2\};\max\{w_1,w_2\}]\!].
\end{equation}
\end{lem}

\begin{proof}
The statement is straightforward for CSS codes by inspecting their parity-check matrices. For general stabilizer codes, suppose $G$ defines a $[\![n,k,d;w]\!]$ code.
Since $|G|=|G_1||G_2|=2^{(n_1-k_1)+(n_2-k_2)}$, we have
\begin{equation}
k=(n_1+n_2)-\log_2|G|=k_1+k_2.
\end{equation}

By definition, we have
\begin{align}
d=&\min\{\wt(E):  E\in\hat{N}(G)-\hat{G}\},\\
d_i=&\min\{\wt(E):  E\in\hat{N}(G_i)-\hat{G_i}\},\quad i=1,2.
\end{align}
We have $N(G)=N(G_1)\otimes N(G_2)$.
Indeed, if $P_1\otimes P_2\in N(G)$, then it commutes with $g_1\otimes I$ and $I\otimes g_2$, for all $g_i\in G_i$, hence $P_i\in N(G_i)$; the converse is immediate.
Hence $\hat{N}(G)=\hat{N}(G_1)\otimes \hat{N}(G_2)$.

For every $E\in\hat{N}(G)-\hat{G}$, write $E=E_1\otimes E_2$, then $E_i\in\hat{N}(G_i)$. At least one of $E_i\notin\hat{G}_i$, otherwise by $\hat{G}=\hat{G}_1\otimes\hat{G}_2$ we know $E_1\otimes E_2\in \hat{G}$. Thus, this $E_i\in\hat{N}(G_i)-\hat{G}_i$, implying $\wt(E)\ge\wt(E_i)\ge d_i$. Therefore, we have
\begin{equation}
d\ge\min\{d_1,d_2\}.
\end{equation}

For $i=1,2$, there exists an $E_i\in\hat{N}(G_i)-\hat{G_i}$ with $\wt(E_i)=d_i$ (If $\hat{N}(G_i)-\hat{G_i}=\emptyset$, i.e., $d_i=\infty$, we simply pick no element). Let $E_1'=E_1\otimes I$ and $E_2'=I\otimes E_2$, satisfying $\wt(E_i')=\wt(E_i)$. Both $E_i'\in\hat{N}(G)-\hat{G}$. Therefore, we have
\begin{equation}
d\le\min\{d_1,d_2\}.
\end{equation}

Recall that by definition,
\begin{align}
w=&\min_{S\subset G,\langle S\rangle=G}\max_{s\in S}\wt(s),\\
w_i=&\min_{S\subset G_i,\langle S\rangle=G_i}\max_{s\in S}\wt(s),\quad i=1,2.
\end{align}

Let $S_i$ be a generating set of $G_i$ achieving $w_i$. Then $S\coloneqq \{s_1\otimes I: \ s_1\in S_1\}\cup\{I\otimes s_2: \ s_2\in S_2\}$
generates $G$ and satisfies $\max_{s\in S}\wt(s)=\max\{w_1,w_2\}$, so
\begin{equation}
w\le \max\{w_1,w_2\}.
\end{equation}

Take any generating set $S\subset G$ and write each $s\in S$ as $s=s_1\otimes s_2$ with $s_i\in G_i$. The projection maps
\begin{equation}
\varphi_i:G\to G_i,\quad s_1\otimes s_2\mapsto s_i,
\end{equation}
are surjective homomorphisms, hence
\begin{equation}
\langle \varphi_i(S)\rangle=\varphi_i(\langle S\rangle)=\varphi_i(G)=G_i.
\end{equation}
By the definition of $w_i$ we have $\max_{t\in \varphi_i(S)}\wt(t)\ge w_i$.
Consequently,
\begin{align}
\begin{aligned}
    \max_{s\in S}\wt(s)\ge&\max\big\{\max_{t\in \varphi_1(S)}\wt(t),\max_{t\in \varphi_2(S)}\wt(t)\big\}\\
\ge&\max\{w_1,w_2\}.
\end{aligned}
\end{align}
Minimizing over $S$ gives
\begin{equation}
w\ge \max\{w_1,w_2\}.
\end{equation}
Collecting the above parts gives the claimed parameters.
\end{proof}

\begin{lem}\label{lemma:clean_weight1}
Suppose $G$ is a stabilizer group that defines an $[\![n,k,d;w]\!]$ stabilizer code. If $\#\{g\in G: \wt(g)=1\}=t$, then there exists a stabilizer group $G'$ that defines an $[\![n-t,k,d;w]\!]$ stabilizer code, and $\#\{g\in G': \wt(g)=1\}=0$.
\end{lem}
\begin{proof}
Let $\{g_1,\cdots,g_{n-k}\}$ be a set of weight-optimal generators for this code, with $\wt(g_1)\le\cdots\le\wt(g_{n-k})$. Since all weight-1 elements in a stabilizer group are independent, we know $g_1,\cdots,g_{t}$ are exactly the $t$ elements in $G$ with weight 1.
Let $J\coloneqq \supp(g_1)\cup\cdots\cup\supp(g_{t})$, we have $\abs{J}=t$.
Denote
\begin{equation}
P_j=\operatorname{sgn}(g_j)g_j|_{\supp(g_j)},\quad j=1,\cdots,t,
\end{equation}
where $\operatorname{sgn}(g_j)$ is the global phase of $g_j$.
For $i=t+1,\cdots,n-k$, we must have $J\bigcap\supp(g_i)=\emptyset$, because otherwise we can replace $g_i$ by $g_i\prod_{j=1}^{t}g_j^{\mathbf{1}_{\supp(g_j)\in\supp(g_i)}}$ in $\{g_1,\cdots,g_{n-k}\}$, which has weight $<\wt(g_i)$, contrary to the weight-optimal property of $\{g_1,\cdots,g_{n-k}\}$.

Discarding the tensor factor $I$ on $J$ for $\{g_{t+1},\cdots,g_{n-k}\}$, we define
\begin{equation}
g'_{i}\coloneqq \operatorname{sgn}(g_i)g_{i}|_{[n]-J},\quad i=t+1,\cdots,n-k.
\end{equation}
$\{g_{t+1}',\cdots,g_{n-k}'\}$ are independent.
Consider the $[\![n-t,k]\!]$ stabilizer code given by $G'\coloneqq \langle g'_{t+1},\cdots,g'_{n-k}\rangle$.
We have
\begin{equation}
G=G'\otimes\Big(\bigotimes_{j=1}^{t}\langle P_j\rangle\Big).
\end{equation}
There are no weight-1 elements in $G'$, because otherwise we can find $>t$ weight-1 elements in $G$.
Note that $\bigotimes_{j=1}^{t}\langle P_j\rangle$ has parameters $[\![t,0,\infty;1]\!]$.
By Lemma~\ref{lemma:tensor_product_parameters}, we know
\begin{equation}
d=\text{distance of } G=\min\{\text{distance of } G',\infty\},
\end{equation}
implying $\text{distance of } G'=d$.
Again by Lemma~\ref{lemma:tensor_product_parameters}, we know
\begin{equation}
w=\mathrm{W}(G)=\max\big\{\w(G'),1\big\},
\end{equation}
implying $\operatorname{W}(G')=w$.
Therefore, the stabilizer code defined by $G'$ has parameters $[\![n-t,k,d;w]\!]$, saving $t$ qubits while achieving the same $k,d,w$, compared to the code $G$.
\end{proof}

\section{Weight-optimal generating set}\label{app:Existence_weight_optimal}

\begin{prop}
For every stabilizer group, there exists a weight-optimal generating set.
\end{prop}
\begin{proof}
Let $G$ be a stabilizer group with $\abs{G}=2^r$.
If $G=\{I\}$, the empty generating set is weight-optimal, so assume $r\ge1$.
We take $g_1\in G\setminus\{I\}$ such that
\begin{equation}
\wt(g_1)=\min_{g\in G\setminus\{I\}}\wt(g).
\end{equation}
For $j=2,\cdots,r$, we take $g_j\in G-\langle g_1,\cdots,g_{j-1}\rangle$ such that
\begin{equation}
\wt(g_j)=\min_{g\in G-\langle g_1,\cdots,g_{j-1}\rangle}\wt(g).
\end{equation}
By construction, $S\coloneqq \{g_1,\cdots,g_r\}$ satisfies $\abs{\langle S\rangle}=2^r$, hence $\langle S\rangle=G$.
For $j=2,\cdots,r$, if an element $h\in G$ satisfies $\wt(h)<\wt(g_{j})$, by the choice of $g_{j}$, we must have $h\in\langle g_1,\cdots,g_{j-1}\rangle$.

Now we show that $S$ is weight-optimal.
Suppose $G$ has another set of generators $S'=\{h_1,\cdots,h_r\}$, with $\wt(h_1)\le\cdots\le\wt(h_r)$.
By $\abs{S'}=r$ we know elements in $S'$ are independent.
By the choice of $g_1$, we have $\wt(g_1)\le\wt(h_1)$. For $i=2,\cdots,r$, we must have $\wt(g_i)\le\wt(h_i)$, because $\wt(h_i)<\wt(g_i)$ would imply $\{h_1,\cdots,h_i\}\subset\langle g_1,\cdots,g_{i-1}\rangle$, contradicting the independence of $\{h_1,\cdots,h_i\}$.
\end{proof}

\section{Computational complexity of calculating optimal weight}\label{app:complexity}

We represent the $n$-qubit Pauli group in the standard binary symplectic form:
each Pauli operator is encoded by a vector in $\mathbb{F}_2^{2n}$.
An instance of \MW is specified by $n-k$ commuting Pauli generators, hence has input length $O(n(n-k))$.
Throughout this section, ``polynomial time'' refers to polynomial in the input length.

\begin{defn}[Minimum-weight stabilizer generators (\MW)]\label{def:mw-sg}
~\\
\begin{itemize}
\item \emph{Instance:} An $n$-qubit stabilizer group $G$ with $|G|=2^{n-k}$, given by generators
$\{P_1,\cdots,P_{n-k}\}\subset \mathcal{P}_n$, and an integer $t\ge 0$.
\item \emph{Question:} Does there exist a generating set $S\subset G$ such that $\max_{s\in S}\wt(s)\le t$?
\end{itemize}
\end{defn}

\begin{lem}[Bounded-size certificates]\label{lem:cert-size}
Let $G\cong \mathbb{F}_2^{\,r}$ with $r=n-k$. If $T\subset G$ generates $G$, then $T$ contains a subset
$S\subset T$ with $|S|=r$ that still generates $G$. Moreover,
$\max_{s\in S}\wt(s)\le \max_{t\in T}\wt(t)$.
\end{lem}

\begin{proof}
Viewing $G$ as an $r$-dimensional vector space over $\mathbb{F}_2$,
any generating set is a spanning set and hence contains a basis of size $r$.
Taking a subset cannot increase the maximum weight.
\end{proof}

\begin{thm}\label{thm:mwsg-npc}
\MW is \NPcomplete.
\end{thm}

\begin{proof}
\emph{Step 1: \MW $\in \NP$.}
By Lemma~\ref{lem:cert-size}, it suffices to consider certificates consisting of exactly $n-k$ elements.
A certificate for a YES-instance $(G,t)$ is a generating set $\{s_1,\cdots,s_{n-k}\}\subset G$
such that $\max_i \wt(s_i)\le t$.
Assume $G=\langle P_1,\cdots,P_{n-k}\rangle$ is given by generators.
A verifier checks in polynomial time that:
\begin{enumerate}
\item $s_i\in \langle P_1,\cdots,P_{n-k}\rangle$ for all $i$;
\item $P_j\in \langle s_1,\cdots,s_{n-k}\rangle$ for all $j$;
\item $\max_i \wt(s_i)\le t$.
\end{enumerate}
Items 1,2 imply $\langle s_1,\cdots,s_{n-k}\rangle = \langle P_1,\cdots,P_{n-k}\rangle = G$.
Both membership tests can be performed efficiently using the binary symplectic representation
(e.g.~\cite{gottesman2016surviving}) by solving linear systems over $\mathbb{F}_2$.

\medskip
\emph{Step 2: \MW is \NPhard.}
Ref.~\cite{Berlekamp1978inherent} proved that \MLD is \NPcomplete via a reduction from \ThreeDM.
We now give a polynomial-time reduction chain
\begin{equation}
\MLD \leqp \SBPFtwo \leqp \MW .
\end{equation}

\begin{defn}[Maximum-likelihood decoding (\MLD)~\cite{Berlekamp1978inherent,Vardy1997intractability}]\label{def:mld}
\par\noindent
\begin{itemize}
\item \emph{Instance:} A binary $m\times n$ matrix $H$ of rank $m$, a vector $\s\in\mathbb{F}_2^m$,
and an integer $t\ge 0$.
\item \emph{Question:} Is there a vector $\e\in\mathbb{F}_2^n$ such that $H\e=\s$ and $\wt(\e)\le t$?
\end{itemize}
\end{defn}

\begin{defn}[Shortest basis problem on $\mathbb{F}_2$ ($\SBPFtwo)$]\label{def:sbp}
For any ambient dimension $N\ge 1$:
\begin{itemize}
\item \emph{Instance:} A linear subspace $V\subset \mathbb{F}_2^N$ of dimension $r$, given by a basis
$\{\mathbf{v}_1,\cdots,\mathbf{v}_r\}\subset \mathbb{F}_2^N$, and an integer $t\ge 0$.
\item \emph{Question:} Does $V$ admit a basis $\{\mathbf{b}_1,\cdots,\mathbf{b}_r\}$ such that
$\max_i \wt(\mathbf{b}_i)\le t$?
\end{itemize}
\end{defn}

\noindent\emph{Step 2.1: $\MLD \leqp \SBPFtwo$.}
Let $(H,\s,t)$ be an instance of \MLD with $H\in\mathbb{F}_2^{m'\times n}$ of rank $m'$ and $\s\in\mathbb{F}_2^{m'}$.
Let $m=n-m'$. Compute a basis $\{\mathbf{a}_1,\cdots,\mathbf{a}_m\}$ for $\ker(H)$ and find $\x\in\mathbb{F}_2^n$ such that $H\x=\s$.
Define
\begin{equation}
B=\left\{
\begin{bmatrix}\mathbf{a}_1\\[2pt]0\end{bmatrix},\cdots,
\begin{bmatrix}\mathbf{a}_m\\[2pt]0\end{bmatrix},
\begin{bmatrix}\x\\[2pt]1^n\end{bmatrix}
\right\}\subset \mathbb{F}_2^{2n},
\qquad
V\coloneqq \operatorname{span}_{\mathbb{F}_2}(B),
\end{equation}
and output the $\SBPFtwo$ instance $(V,n+t)$.

If $(H,\s,t)$ is a YES-instance, there exists $\e\in\mathbb{F}_2^n$ such that $H\e=\s$ and $\wt(\e)\le t$.
Then $H(\e+\x)=0$, hence $\e+\x=\sum_{i=1}^m c_i\mathbf{a}_i$ for some $c_i\in\mathbb{F}_2$. Therefore,
\begin{equation}
\left\{
\begin{bmatrix}\mathbf{a}_1\\[2pt]0\end{bmatrix},\cdots,
\begin{bmatrix}\mathbf{a}_m\\[2pt]0\end{bmatrix},
\begin{bmatrix}\e\\[2pt]1^n\end{bmatrix}
\right\}
\text{ is a basis of }V,
\qquad
\wt\!\left(\begin{bmatrix}\e\\ 1^n\end{bmatrix}\right)\le n+t,
\end{equation}
so $(V,n+t)$ is a YES-instance.

If $(H,\s,t)$ is a NO-instance, then for every $\e\in\mathbb{F}_2^n$ with $H\e=\s$ we have $\wt(\e)>t$.
Let $\{\mathbf{w}_1,\cdots,\mathbf{w}_{m+1}\}$ be any basis of $V$.
Since all vectors in $\operatorname{span}\{\big[\!\begin{smallmatrix}\mathbf{a}_i\\0\end{smallmatrix}\!\big]\}$ have the last $n$ coordinates equal to $0$,
at least one basis vector must have the last $n$ coordinates equal to $1^n$; write it as
\begin{equation}
\mathbf{w}_1=\sum_{i=1}^m c_i\begin{bmatrix}\mathbf{a}_i\\[2pt]0\end{bmatrix}+\begin{bmatrix}\x\\[2pt]1^n\end{bmatrix}
=\begin{bmatrix}\x+\sum_{i=1}^m c_i\mathbf{a}_i\\[2pt]1^n\end{bmatrix}.
\end{equation}
Then $H\!\left(\x+\sum_{i=1}^m c_i\mathbf{a}_i\right)=H\x=\s$, so the top block has weight $>t$, which implies $\wt(\mathbf{w}_1)>n+t$.
Hence $(V,n+t)$ is a NO-instance.

\medskip
\noindent\emph{Step 2.2: $\SBPFtwo \leqp \MW$.}
Let $(V,t)$ be an instance of \SBPFtwo with $V\subset\mathbb{F}_2^n$ of dimension $r$ given by a basis $\{\mathbf{v}_1,\cdots,\mathbf{v}_r\}$.
For $\mathbf{u}=(u_1,\cdots,u_n)\in\mathbb{F}_2^n$, define the $Z$-type Pauli operator
\begin{equation}
Z(\mathbf{u}) \coloneqq Z^{u_1}\otimes\cdots\otimes Z^{u_n}.
\end{equation}
Let $G\coloneqq \{Z(\mathbf{u}) : \mathbf{u}\in V\}$. Then $G$ is an abelian stabilizer group generated by $\{Z(\mathbf{v}_1),\cdots,Z(\mathbf{v}_r)\}$, and $\wt(Z(\mathbf{u}))=\wt(\mathbf{u})$.
Moreover, a set $S\subset G$ generates $G$ if and only if the corresponding vectors span $V$; by Lemma~\ref{lem:cert-size},
we may restrict to generating sets of size $r$ without increasing the maximum weight.
Therefore, $(V,t)$ is a YES-instance of $\SBPFtwo$ if and only if $(G,t)$ is a YES-instance of \MW.

This completes the proof that \MW is \NPhard, and hence \NPcomplete.
\end{proof}

\section{Proof of the basic bound in Theorem~\ref{thm:w_LBby_nk} in the main text}\label{app:proof_nk_bound}

By definition, there exists a stabilizer group $G$ with parameters $[\![n,k,d;w]\!]$, such that $w=\mathrm{W}_{\!\mathrm{opt}}(n,k,d)$.
Denote
\begin{equation}
t=\#\{g\in G:\wt(g)=1\},
\end{equation}
by Lemma~\ref{lemma:clean_weight1}, there exists a stabilizer group $G'$ with parameters $[\![n-t,k,d;w]\!]$, and there are no weight-1 elements in $G'$.
Take $\{h_1,\cdots,h_{n-t-k}\}$ as a weight-optimal generating set for $G'$, satisfying $\wt(h_1)\le\cdots\le\wt(h_{n-t-k})=w$.

Now we show that each qubit is covered by at least two elements in $\{h_1,\cdots,h_{n-t-k}\}$.
If there exists a qubit $q\in[n-t]$, such that none of $h_1,\cdots,h_{n-t-k}$ covers $q$, then single qubit Paulis on $q$ are logical error for the code $G'$, implying $d=1$, contrary to our assumption $d\ge2$.
If there exists a qubit $q\in[n-t]$, such that only one element in $\{h_1,\cdots,h_{n-t-k}\}$ covers $q$, say $h_i$, then $h_i|_{q}\otimes I_{[n-t]-\{q\}}$ is a logical error for $G'$, implying $d=1$, a contradiction.

Notice that
\begin{equation}
\sum_{i=1}^{n-t-k}\wt(h_i)=\sum_{q=1}^{n-t}\#\{i\in[n-t-k]:q\in\supp(h_i)\}
\end{equation}
Also we have
\begin{equation}
\sum_{i=1}^{n-t-k}\wt(h_i)\le (n-t-k)w
\end{equation}
and
\begin{equation}
\sum_{q=1}^{n-t}\#\{i\in[n-t-k]:q\in\supp(h_i)\}\ge2(n-t),
\end{equation}
Therefore, we have $(n-t-k)w\ge2(n-t)$, implying
\begin{equation}
\mathrm{W}_{\!\mathrm{opt}}(n,k,d)=w\ge\frac{2(n-t)}{n-t-k}\ge\frac{2n}{n-k}.
\end{equation}

\section{Stabilizer codes with  weight $\le3$}\label{app:wle3}
In this section, we want to derive the behavior of codes with weight $3$. Our main result is the following theorem.
\begin{thm}\label{thm:wle3}
Suppose $k\ge 1$ and $d\ge 2$.
If there exists an $[\![n,k,d;3]\!]$ stabilizer code, then $d=2$ and $k/n\le 1/4$.
\end{thm}
We first prove that $d\le2$, then derive the rate bound by counting the incidences between generators and qubits.

\begin{lem}\label{lem:wle3_distance}
If a stabilizer code with $k\ge1$ has a generating set of weight at most $3$, then $d\le2$.
\end{lem}
\begin{proof}
Suppose to the contrary that $d\ge3$, and choose a counterexample with stabilizer group $G$ on the smallest possible number of qubits.
Among all independent weight-$\le3$ generating sets of $G$, choose $\mc{B}=(g_1,\cdots,g_m)$ minimizing $\sum_i\wt(g_i)$, where $m=n-k<n$.

The minimality assumptions impose two restrictions that we will use throughout the proof.
If $a$ independent stabilizers of weight at most $3$ are supported on a set $A$ with $|A|\le a$, isotropy gives $a\le|A|$, so equality holds.
Their restrictions form a maximal abelian Pauli subgroup on $A$.
Extend them to a weight-$\le3$ basis and clear the restriction of every remaining basis element on $A$ by multiplying stabilizers supported on $A$.
This does not increase its weight and splits off a stabilizer-state factor on $A$.
The same maximal subgroup clears the restriction to $A$ of every normalizer element, so the complementary factor has the same logical quotient and distance.
It is therefore a smaller counterexample with the same $k$ and $d$.
Thus, $a$ independent stabilizers of weight at most $3$ cannot be supported on $a$ or fewer qubits; we denote this restriction by (F).
In particular, $G$ has no weight-$1$ stabilizer.
Moreover, on every qubit the local restrictions of $G$ span the two-dimensional single-qubit Pauli space: otherwise a nonidentity weight-$1$ Pauli would commute with $G$ and, since $d\ge3$, would have to be a weight-$1$ stabilizer.

For two rows $g,h\in\mc{B}$, let $c$ count common qubits on which their local Paulis coincide and let $a'$ count those on which they differ.
Commutation makes $a'$ even.
Whenever $\wt(gh)\le3$, replacing either row by $gh$ and using minimality gives
\begin{equation}\label{eq:w3_pair_minimality}
\wt(gh)\ge\max\{\wt(g),\wt(h)\},\quad \wt(gh)=\wt(g)+\wt(h)-2c-a'.
\end{equation}
Thus two rows cannot coincide on two qubits, nor can they coincide on one qubit and mismatch on another, since either case gives $\wt(gh)<\max\{\wt(g),\wt(h)\}\le3$.
In particular, a mismatching pair mismatches on exactly two qubits and coincides nowhere.
We denote this restriction on overlaps by (O).

We next show that the rows in $\mc{B}$ use at most two Pauli labels on each qubit.
Otherwise choose three rows $A,B,C$ carrying $X,Y,Z$, respectively, on one qubit.
All three have weight $3$.
Indeed, two weight-$2$ rows would give two independent stabilizers on the same two qubits, violating (F).
If exactly one had weight $2$, commutation and (O) would make $ABC$ a nonidentity Pauli of weight at most $2$, which could replace a weight-$3$ row and lower the total weight.
Up to qubit relabeling and single-qubit Clifford gates, (O) leaves only
\begin{align}
\mathrm{(I)}\quad A&=X_1X_2X_3,\quad B=Z_1Z_2X_4,\quad C=Y_1Y_2X_5,\label{eq:w3_three_axes_I}\\
\mathrm{(II)}\quad A&=X_1X_2X_3,\quad B=Z_1Z_2X_4,\quad C=Y_1Z_3Z_4.\label{eq:w3_three_axes_II}
\end{align}
To see this, label the second mismatch of $A$ and $B$ as qubit $2$.
The second mismatches of $C$ with $A$ and $B$ either both occur at qubit $2$, giving (I), or occur at the remaining qubit of each row, giving (II).

Pattern (I) reduces to pattern (II).
The Pauli $L=X_1X_2$ commutes with $A,B,C$.
If it commuted with every stabilizer, then $L$ would be a stabilizer and $LA=X_3$ would be a forbidden weight-$1$ stabilizer.
Hence some row $D$ anticommutes with $L$.
After exchanging qubits $1,2$ and rows $B,C$ if necessary, take $D|_1=Z$.
Rule (O) excludes $D|_2=X$; commutation with $A$ and $C$ then forces $D$ to mismatch them on qubits $3$ and $5$, respectively.
Consequently, $A,C,D$ have pattern (II), up to local relabeling.

It remains to exclude (II).
Now $L=X_3X_4$ commutes with $A,B,C$.
If it commuted with all of $G$, then $L$ would be a stabilizer.
It is independent of $A,B,C$, because their only nonempty product that is trivial on qubit $1$ is $ABC$, which acts nontrivially on qubit $2$.
These four stabilizers on qubits $1,\cdots,4$ would then violate (F).
Therefore some row $D$ anticommutes with $L$.
After exchanging $A$ with $B$, relabeling qubits $3,4$, and changing local axes if necessary, assume that $D$ mismatches at qubit $3$.
If $D|_3=Y$, then either $D$ is supported on the first four qubits, in which case $A,B,C,D$ violate (F), or commutation and (O) force
\begin{equation}
D=Z_1Y_3X_5
\end{equation}
after relabeling its new qubit.
If $D|_3=Z$, then it coincides with $C$ at qubit $3$; rule (O) forbids it from touching qubits $1,4$, and commutation with $A,B$ forces
\begin{equation}
D=Z_2Z_3X_5,
\end{equation}
where a new qubit is again required by (F).

The Pauli $X_5$ cannot commute with every stabilizer, because it would then be a weight-$1$ stabilizer.
Choose a row $E$ that anticommutes with it, so $E|_5\in\{Y,Z\}$.
Rule (O) says that $E$ has one further mismatch with $D$, no coinciding overlap with $D$, and no support on the other old support position of $D$.
Commutation with $A,B,C$ determines the remaining factor of $E$:
\begin{equation}\label{eq:w3_E_table}
\begin{array}{c|c|c} D & \text{second mismatch with }D & \text{remaining factor of }E\\ \hline Z_1Y_3X_5 & E|_1=X & E|_4=Y\\ & E|_1=Y & E|_2=Y\\ & E|_3=X & E|_4=X\\ & E|_3=Z & E|_2=Z\\ \hline Z_2Z_3X_5 & E|_2=X & E|_4=Z\\ & E|_2=Y & E|_1=Y\\ & E|_3=X & E|_4=X\\ & E|_3=Y & E|_1=Z \end{array}.
\end{equation}
These are all possibilities for the second mismatch with $D$ and its Pauli label.
Hence $E$ is supported on qubits $1,\cdots,5$.
Since $A,B,C,D,E$ are distinct rows of the basis, they are independent, and (F) gives a contradiction.
Thus exactly two Pauli labels occur among the basis rows on every qubit.

Apply single-qubit Clifford gates to call these labels $X$ and $Z$.
Form a graph $\Gamma$ on the rows of $\mc{B}$, joining two rows when they mismatch.
We claim that $\Gamma$ is bipartite.
Otherwise take a shortest odd cycle of length $\ell$; it is chordless.
Every row on it has weight $3$, since the two neighbours of a weight-$2$ row would both mismatch it on both support qubits and hence coincide with each other twice, contrary to (O).

Every qubit in the union of the cycle supports occurs in at least three cycle rows.
If it occurred once, the two neighbouring rows would coincide on the other two support qubits.
The same argument applies if it occurred twice with the same label, because the two rows carrying that label cannot be adjacent.
If it occurred twice with different labels, those rows must be consecutive and, by (O), mismatch at a second qubit.
Their other cycle neighbours then mismatch each other at that second qubit.
For $\ell\ge5$, this gives a chord.
For $\ell=3$, the common neighbour would need a label different from both $X$ and $Z$ at the same qubit, which is impossible.
Thus the $\ell$ independent weight-$3$ rows on the cycle are supported on at most $\ell$ qubits, contradicting (F).
Hence $\Gamma$ is bipartite.

Let its two parts be $\mc{B}_X$ and $\mc{B}_Z$.
Both parts meet every qubit: rows within one part never mismatch, so one part alone would give a one-dimensional local span.
Orient the local Clifford frame at each qubit so that all incident rows of $\mc{B}_X$ carry $X$.
Every incident row of $\mc{B}_Z$ must then carry $Z$.
Thus the transformed basis is CSS.

It remains to rule out this CSS counterexample.
Rows of the same type intersect in at most one qubit by (O).
Every qubit belongs to at least one and at most two rows of each type.
Indeed, three $Z$-rows through one qubit would have pairwise disjoint remaining supports.
An $X$-row through that qubit would then need a distinct second intersection with each of them, giving weight at least $4$.
The same argument applies with $X$ and $Z$ interchanged.

Let $r_2,r_3$ be the numbers of weight-$2$ and weight-$3$ rows, and let $t$ count unordered intersecting pairs of rows of the same type.
Counting incidences gives
\begin{equation}\label{eq:w3_css_incidence}
2r_2+3r_3=2n+t.
\end{equation}
To show that $t\ge r_3$, we assign to each weight-$3$ row $A$ two distinct ordered pairs of intersecting rows of the same type.
Such a row has a same-type neighbour: if it had none and $\supp(A)=\{i,j,s\}$, the opposite-type pair Paulis on $\{i,j\}$ and $\{i,s\}$ would both be stabilizers, and together with $A$ would violate (F).
If $A$ has two same-type neighbours $B_1,B_2$, assign $(A,B_1)$ and $(A,B_2)$.

Otherwise $A$ has a unique same-type neighbour $B$, meeting it, say, at $s$.
The opposite-type pair Pauli $C$ on the remaining sites $i,j$ commutes with all of $G$ and hence is a weight-$2$ stabilizer.
In fact $C$ is a row of $\mc{B}$.
Its basis expansion uses only rows of its CSS type, and minimal total weight forces all of them to have weight $2$.
Any such row through $i$ must use $j$ or $s$ to commute with $A$; it cannot use $s$, because $B$ contains $s$ but neither $i$ nor $j$.
Hence the only possible weight-$2$ row through $i$ is $C$ itself, and parity at $i$ in the expansion forces $C\in\mc{B}$.
Finally, $C$ has a same-type neighbour $D$; otherwise $C$ and the complementary-type pair Pauli on $\{i,j\}$ would be two independent stabilizers on two qubits, violating (F).
Assign $(A,B)$ and $(C,D)$.

The assigned ordered pairs are all distinct.
A pair whose first row has weight $3$ determines $A$ directly.
A pair $(C,D)$ also determines $A$, since two weight-$3$ rows of the same type containing $\supp(C)$ would intersect twice.
Since the $t$ unordered pairs give $2t$ ordered pairs, $2r_3\le2t$.
Eq.~\eqref{eq:w3_css_incidence} therefore yields
\begin{equation}
2n=2r_2+3r_3-t\le2(r_2+r_3)=2m,
\end{equation}
contrary to $m=n-k<n$.
\end{proof}

\begin{lem}\label{lem:wle3_remove_weight_one}
If the stabilizer group $G$ of an $[\![n,k,d;3]\!]$ code contains a weight-$1$ element, then the code factorizes as a one-qubit stabilizer state tensored with an $[\![n-1,k,d;3]\!]$ code.
\end{lem}
\begin{proof}
Set $m\coloneqq n-k$, let the weight-$1$ stabilizer be $g=P_q$, and begin with a basis of $G$ whose elements have weight at most three.
The images of this basis span the quotient $G/\langle g\rangle$; select a quotient basis from these images and let $h_1,\cdots,h_{m-1}$ be the corresponding original basis elements.
Then $g,h_1,\cdots,h_{m-1}$ form a basis of $G$.
Because each $h_i$ commutes with $g$, its local Pauli on $q$ is either $I$ or $P$.
In the latter case, replace $h_i$ by $h_i g$ to clear its action on $q$ without increasing its weight.
The resulting basis gives a factorization $G=G_0\otimes\langle g\rangle$, where $G_0$ acts on the other $n-1$ qubits and $\langle g\rangle$ defines an $[\![1,0,\infty;1]\!]$ stabilizer state.
By Lemma~\ref{lemma:tensor_product_parameters}, $G_0$ retains $k$ logical qubits and distance $d$, while $3=\w(G)=\max\{1,\w(G_0)\}$.
Thus, $\w(G_0)=3$, which proves the claimed parameters.
\end{proof}

We now prove the rate bound.
By applying Lemma~\ref{lem:wle3_remove_weight_one} repeatedly, we remove all fixed one-qubit stabilizer states and obtain a code on $n_0\le n$ qubits, which we call the core.
The core still has parameters $[\![n_0,k,d;3]\!]$, and its stabilizer group contains no weight-$1$ element.
Let $m\coloneqq n_0-k$ and choose an independent generating set $S=\{g_1,\cdots,g_m\}$ in which every generator has weight at most three.

\begin{lem}\label{lem:wle3_local_rank}
On every core qubit $q$, the restrictions $\{g_i|_q\}_{i=1}^m$ span the two-dimensional single-qubit Pauli space modulo phases.
In particular, if
\begin{equation}
\delta_q\coloneqq\bigl|\{i:q\in\supp(g_i)\}\bigr|,
\end{equation}
then $\delta_q\ge2$.
\end{lem}
\begin{proof}
If the local span had dimension at most one, some nonidentity Pauli supported only on $q$ would commute with every generator.
Since $d\ge2$, this Pauli would have to be a stabilizer, contrary to the definition of the core.
\end{proof}

\begin{lem}\label{lem:wle3_high_degree}
Every weight-$3$ generator $g_i$ contains a qubit $q$ with $\delta_q\ge3$.
\end{lem}
\begin{proof}
Suppose instead that all three qubits in $\supp(g_i)$ have degree two.
At each such qubit, Lemma~\ref{lem:wle3_local_rank} implies that the unique other incident generator has a different local Pauli and hence anticommutes locally with $g_i$.
Since $g_i$ commutes with every $g_j$, the number of their local anticommutations is even for each $j\ne i$.
Summing over $j\ne i$ therefore gives an even number, but counting over the three qubits gives three, a contradiction.
\end{proof}

\begin{lem}\label{lem:wle3_incidence_bound}
The core satisfies $n_0\ge4k$.
\end{lem}
\begin{proof}
Let $a$ and $b$ be the numbers of weight-$2$ and weight-$3$ generators in $S$, respectively, so $m=a+b$.
By Lemma~\ref{lem:wle3_local_rank},
\begin{equation}
E\coloneqq\sum_q(\delta_q-2)=2a+3b-2n_0=2m+b-2n_0\ge0.
\end{equation}
Let $H\coloneqq\sum_{q:\,\delta_q\ge3}\delta_q$ count incidences on qubits of degree at least three.
Lemma~\ref{lem:wle3_high_degree} gives $b\le H$.
Moreover, $\delta\le3(\delta-2)$ for every integer $\delta\ge3$, and hence $H\le3E$.
Consequently $E\ge b/3$, so
\begin{equation}
2n_0=2m+b-E\le2m+\frac{2b}{3}\le\frac{8m}{3}.
\end{equation}
Using $m=n_0-k$ gives $4k\le n_0$.
\end{proof}

\begin{lem}\label{lem:wle3_tight_block}
The rate bound in Lemma~\ref{lem:wle3_incidence_bound} is attainable.
\end{lem}
\begin{proof}
The stabilizer group $\langle XXXI,IYYY,ZIZZ\rangle$ defines an $[\![4,1,2;3]\!]$ code with $k/n=1/4$.
\end{proof}

\begin{proof}[Proof of Theorem~\ref{thm:wle3}]
Lemma~\ref{lem:wle3_distance} and the assumption $d\ge2$ give $d=2$.
For the rate, Lemma~\ref{lem:wle3_incidence_bound} gives $4k\le n_0\le n$, and therefore $k/n\le1/4$.
\end{proof}

\section{Check-weight bounds for distance at least three}\label{app:analytic_rate_bounds}

We prove the remaining bounds in Theorem~\ref{thm:rate_weight_main} of the main text.
Let $\hat{G}$ denote the image, modulo Pauli phases, of the stabilizer group $G$ of an $[\![n,k,d]\!]$ code.
We identify elements of $G$ with their images in $\hat{G}$ whenever taking spans, dimensions, or bases.
Then $\hat{G}$ is an $m$-dimensional vector space over $\mbb{F}_2$, where
\begin{equation}
m=n-k.
\end{equation}
For $j\ge1$, define
\begin{equation}\label{eq:Lj_definition}
L_j(G)\coloneqq\operatorname{span}_{\mbb{F}_2}\{\hat{g}:g\in G,\ \wt(g)\le j\}.
\end{equation}

\subsection{Generating bases and degree-two columns}

For an independent generating set
\begin{equation}
\mc{B}=(g_1,\cdots,g_m),
\end{equation}
let $r_q$ denote the column degree at qubit $q$, and let $I$ count the incidences between generators and qubits:
\begin{equation}\label{eq:column_degree_incidence}
r_q\coloneqq\#\{i:q\in\supp(g_i)\},\quad I\coloneqq\sum_{i=1}^{m}\wt(g_i)=\sum_{q=1}^{n}r_q.
\end{equation}

\begin{lem}\label{lem:active_adapted_basis}
Suppose $G$ defines an $[\![n,k,d;w]\!]$ code with $k\ge1$ and $d\ge2$.
\begin{enumerate}[label=(\roman*)]
\item Let $h\coloneqq\dim L_1(G)$.
The weight-$1$ stabilizers split off as $h$ one-qubit stabilizer-state factors.
The remaining active core has parameters $[\![n-h,k,d;w]\!]$ and contains no weight-$1$ stabilizer.
\item On the active core, the local restrictions of $G$ at every qubit span the two-dimensional single-qubit Pauli space.
Consequently, every column in every generating basis has degree at least $2$.
\item Let $L\subset G$ be spanned by stabilizers of weight at most $u$, and write $\ell=\dim L$.
If $G$ admits a basis of weight at most $w$, then it admits a basis whose first $\ell$ rows have weight at most $u$ and whose remaining rows have weight at most $w$.
In particular,
\begin{equation}\label{eq:adapted_incidence_bound}
I\le u\ell+w(m-\ell).
\end{equation}
\end{enumerate}
\end{lem}

\begin{proof}
For (i), suppose $h>0$, since there is nothing to prove when $h=0$.
Distinct weight-$1$ stabilizers have disjoint supports.
Hence the weight-$1$ stabilizers are linearly independent, their number is $h$, and we may choose all of them as the first elements of a basis of $G$.
Commutation implies that the restriction of every remaining basis element to each of these $h$ qubits is either the identity or the same Pauli as the corresponding weight-$1$ stabilizer.
Multiplying by that stabilizer in the latter case clears the restriction without increasing the weight.
This gives $G=G_0\otimes H$, where $H$ is an $h$-qubit product-state stabilizer group and $G_0$ contains no weight-$1$ stabilizer.
Lemma~\ref{lemma:tensor_product_parameters} shows that $G_0$ retains the logical dimension and distance and that $\w(G)=\max\{1,\w(G_0)\}$.
Moreover, $\w(G_0)\ge2$: otherwise, because $G_0$ contains no weight-$1$ stabilizer, it would be trivial and would admit a weight-$1$ logical Pauli, contradicting $k\ge1$ and $d\ge2$.
Hence $\w(G_0)=w$, proving (i).

For (ii), suppose that the local restrictions on some qubit span a space of dimension at most one.
A nonidentity single-qubit Pauli then commutes with all of $G$.
It cannot be a stabilizer on the active core and would therefore be a logical operator of weight one, contradicting $d\ge2$.

For (iii), extract a basis of $L$ from its weight-$\le u$ spanning set.
The images of any weight-$\le w$ basis of $G$ span the quotient $G/L$, so a subset of these images forms a quotient basis.
Adjoining the corresponding representatives to the chosen basis of $L$ gives the required basis.
\end{proof}

Fix an active core and an adapted basis.
Define a graph $\Gamma_2$ on the $m$ generator rows by associating each degree-$2$ column $q$ with an edge joining the two rows that act on it.
This graph has no loops but may have parallel edges.
Define
\begin{equation}\label{eq:degree_two_graph_parameters}
e\coloneqq|E(\Gamma_2)|,\quad v\coloneqq\#\{i:\deg_{\Gamma_2}(i)>0\},\quad b\coloneqq m-v,
\end{equation}
and
\begin{equation}\label{eq:s_and_t_definition}
s\coloneqq\sum_{i:\deg_{\Gamma_2}(i)>0}\bigl(\deg_{\Gamma_2}(i)-1\bigr)=2e-v,\quad t\coloneqq\sum_{q:r_q\ge4}(r_q-3).
\end{equation}

\begin{lem}\label{lem:degree_two_star_spaces}
If $d\ge3$, then the degree-$2$ graph determines a direct sum $K\subset L_2(G)$ of weight-$2$ stabilizer spaces with
\begin{equation}\label{eq:star_space_dimension}
\dim K=s.
\end{equation}
Consequently, $s\le\dim L_2(G)$ and
\begin{equation}\label{eq:e_identity}
2e=m-b+s.
\end{equation}
\end{lem}

\begin{proof}
For a degree-$2$ column $q=(i,j)$, the two local Pauli labels must be linearly independent; otherwise, the local span at $q$ would be one-dimensional, contrary to Lemma~\ref{lem:active_adapted_basis}(ii).
Hence the map from a single-qubit Pauli on $q$ to its two syndrome bits at rows $i$ and $j$ is an isomorphism of $\mbb{F}_2^2$.

For every edge $q$ incident on row $i$, let $E_{i,q}$ be the unique single-qubit Pauli whose syndrome is the unit vector $\mathbf{e}_i$.
After fixing one incident edge $q_0$, the products $E_{i,q}E_{i,q_0}$ over all other incident edges are independent weight-$2$ centralizer elements.
Since $d\ge3$, these elements are stabilizers and span a space $K_i$ of dimension $\deg_{\Gamma_2}(i)-1$.

The spaces $K_i$ form a direct sum.
Indeed, on an edge $q=(i,j)$, the restrictions contributed by $K_i$ and $K_j$ lie along the two independent single-qubit Paulis $E_{i,q}$ and $E_{j,q}$.
Thus, if a sum of elements from the spaces $K_i$ vanishes, each endpoint contributes zero on every edge.
This proves Eq.~\eqref{eq:star_space_dimension}.
Eq.~\eqref{eq:e_identity} follows from $s=2e-v$ and $v=m-b$.
\end{proof}

Let $M$ count the isolated edges of $\Gamma_2$, whose two endpoints both have graph degree one.

\begin{lem}\label{lem:isolated_edge_lower_bound}
The number of isolated edges satisfies
\begin{equation}\label{eq:M_lower_bound}
M\ge e-2s=\frac{m-b-3s}{2}.
\end{equation}
\end{lem}

\begin{proof}
Every nonisolated edge is incident on a vertex of graph degree at least two.
Therefore,
\begin{equation}
e-M\le\sum_{i:\deg(i)\ge2}\deg(i)\le2\sum_{i:\deg(i)\ge2}(\deg(i)-1)\le2s.
\end{equation}
Combining this with Eq.~\eqref{eq:e_identity} proves the result.
\end{proof}

For an isolated edge $q=(i,j)$, the rows $g_i$ and $g_j$ anticommute locally on $q$.
Since they commute globally, their local restrictions must also anticommute on another column $r$, which we call a \emph{mismatch partner} of $q$.
This partner has degree at least three: a second degree-$2$ column on the same endpoints would make $q$ nonisolated.
We choose a degree-$3$ partner whenever one is available and otherwise choose any partner of degree at least four.

\subsection{Stabilizer codes with distance at least three}

\begin{thm}\label{thm:d3_rate_weight_app}
Let $G$ define an $[\![n,k,d;w]\!]$ stabilizer code with $k\ge1$, $d\ge3$, and $w\ge4$.
Then
\begin{equation}\label{eq:d3_rate_weight_app}
12n\le(4w+1)(n-k)-1.
\end{equation}
\end{thm}

\begin{proof}
We first work on the active core and continue to denote its length and stabilizer rank by $n$ and $m$.
Set $\ell\coloneqq\dim L_2(G)$.
Choose an $L_2(G)$-adapted basis by Lemma~\ref{lem:active_adapted_basis}, and set
\begin{equation}\label{eq:A_d3_definition}
A\coloneqq wm-(w-2)\ell-I\ge0.
\end{equation}
Since every column has degree at least two,
\begin{equation}\label{eq:I_degree_decomposition}
I=3n-e+t.
\end{equation}
Let $c\coloneqq\ell-s\ge0$.
Combining Eqs.~\eqref{eq:e_identity}, \eqref{eq:A_d3_definition}, and \eqref{eq:I_degree_decomposition} gives
\begin{equation}\label{eq:Delta_d3_identity}
\Delta\coloneqq(2w+1)m-6n=(2w-5)\ell+b+c+2A+2t.
\end{equation}

We next bound the number $M$ of isolated edges from above.
Suppose that the chosen mismatch partner $r$ of an isolated edge $q$ has degree three.
Choose a nonzero single-qubit Pauli on $r$ that commutes with the third row incident on $r$.
Its syndrome on the other two rows is nonzero and can therefore be matched by a unique single-qubit Pauli on $q$.
The product of these two Paulis is a weight-$2$ centralizer element and hence belongs to $L_2(G)$.
Let $G_3$ count the isolated edges assigned a degree-$3$ partner.
The resulting stabilizers are linearly independent and are also independent of the star space $K$: each acts on its own isolated edge, whereas $K$ vanishes on every isolated edge.
Thus,
\begin{equation}\label{eq:G3_d3_bound}
s+G_3\le\ell.
\end{equation}
Assign every remaining isolated edge to a partner of degree $r\ge4$, and let $\operatorname{occ}(q)$ denote the number of isolated edges assigned to column $q$.
Because isolated edges have pairwise disjoint endpoints, a degree-$r$ column can serve at most $\lfloor r/2\rfloor\le2(r-3)$ such edges.
It follows that
\begin{equation}\label{eq:M_upper_d3}
M\le G_3+2t\le\ell-s+2t.
\end{equation}
Combining Eqs.~\eqref{eq:M_lower_bound} and \eqref{eq:M_upper_d3} yields
\begin{equation}\label{eq:structure_d3}
m\le2\ell+s+b+4t\le3\ell+b+4t.
\end{equation}

By Eq.~\eqref{eq:structure_d3}, we have
\begin{equation}
H\coloneqq3\ell+b+4t-m\ge0.
\end{equation}
For $w\ge4$, Eq.~\eqref{eq:Delta_d3_identity} then gives
\begin{equation}\label{eq:Gamma_d3_decomposition}
\Gamma\coloneqq(4w+1)m-12n=H+(4w-13)\ell+b+2c+4A\ge0.
\end{equation}
We now show that this inequality is strict.
Suppose that $(4w+1)m-12n=0$.
Then equality holds in the preceding inequalities, giving
\begin{equation}\label{eq:d3_equality_conditions}
\ell=s=b=A=0,\quad e=M=\frac{m}{2},\quad t=\frac{m}{4}.
\end{equation}
Thus, the degree-$2$ columns form a perfect matching on the generator rows.
Since we chose a degree-$3$ mismatch partner whenever possible and $G_3\le\ell-s=0$, no matching edge has such a partner.
The remaining $M=m/2=2t$ assignments therefore satisfy
\begin{equation}
M=\sum_q\operatorname{occ}(q)\le\sum_q\left\lfloor\frac{r_q}{2}\right\rfloor\le2\sum_q(r_q-3)=2t,
\end{equation}
where both sums run over columns of degree at least four.
The second inequality is strict for every $r_q\ge5$.
Equality also requires every column of degree at least four to receive an assignment.
Thus every such column has degree four and is assigned exactly two matching edges.
Consequently, the support of every degree-$4$ column is precisely the union of two disjoint matching edges, every row belongs to one degree-$2$ column and one degree-$4$ column, and every remaining column has degree three.

Because $\ell=0$, the three local labels in every degree-$3$ column must be $X,Y,Z$ in some order.
Indeed, Lemma~\ref{lem:active_adapted_basis}(ii) excludes three identical labels.
If instead the labels had the form $P,P,Q$ with $P\ne Q$, the single-qubit error $P$ on this column would have a unit syndrome at the row labeled $Q$.
A single-qubit Pauli on that row's degree-$2$ matching column has the same syndrome.
Their product would then be a weight-$2$ stabilizer, contradicting $\ell=0$.

Fix a generator row.
Summing, modulo two, its pairwise commutation equations with all other rows shows that the total number of local anticommutations involving this row is even.
Its degree-$2$ column contributes one, whereas each degree-$3$ column contributes two.
Its unique degree-$4$ column must therefore contribute an odd number.
On the other hand, this degree-$4$ column consists of two matching pairs, and the two labels within each pair are distinct.
Hence the multiplicities of its four Pauli labels are either $2+2$ or $2+1+1$.
A label of multiplicity $\mu$ has local anticommutation degree $4-\mu$.
In the first case, all four degrees are even; in the second, two are even and two are odd.
The degrees therefore cannot be odd for all four rows, a contradiction.
Thus, the nonnegative integer $(4w+1)m-12n$ is at least one.

Finally, restore the $h$ weight-$1$ factors removed in Lemma~\ref{lem:active_adapted_basis}.
If $n=n_0+h$ and $m=m_0+h$, then
\begin{equation}
(4w+1)m-12n=\bigl[(4w+1)m_0-12n_0\bigr]+(4w-11)h\ge1,
\end{equation}
which proves Eq.~\eqref{eq:d3_rate_weight_app} for the original code.
\end{proof}

\subsection{Stabilizer codes with distance at least four}

\begin{thm}\label{thm:d4_rate_weight_app}
Let $G$ define an $[\![n,k,d;w]\!]$ stabilizer code with $k\ge1$, $d\ge4$, and $w\ge4$.
Then
\begin{equation}\label{eq:d4_rate_weight_app}
3n\le w(n-k).
\end{equation}
\end{thm}

\begin{proof}
Again begin with the active core.
By Lemma~\ref{lem:active_adapted_basis}, choose a basis whose first $\dim L_2(G)$ rows have weight at most two, whose next $\dim L_3(G)-\dim L_2(G)$ rows have weight at most three, and whose remaining $m-\dim L_3(G)$ rows have weight at most $w$.
Counting the row weights gives
\begin{equation}\label{eq:A_d4_definition}
I\le wm-(w-3)\dim L_3(G)-\dim L_2(G).
\end{equation}
The degree-$2$ star space has dimension $s\le\dim L_2(G)$.

We next bound $M$.
As in the preceding proof, each chosen degree-$3$ partner gives a weight-$2$ stabilizer supported on the assigned isolated edge and its partner.
Next, suppose that a degree-$4$ column $r$ is assigned two isolated edges $q_1=(i,j)$ and $q_2=(i',j')$.
For a local Pauli $P$ on $r$, the syndrome isomorphisms on $q_1$ and $q_2$ provide unique local Paulis that cancel the syndrome of $P$ on the two endpoint pairs.
Writing the resulting centralizer element as $F_P$, we obtain a linear map
\begin{equation}
P\longmapsto F_P
\end{equation}
from the two-dimensional single-qubit Pauli space modulo phases to centralizer elements supported on $q_1$, $q_2$, and $r$.
For $P\ne I$, the mismatch condition makes the syndrome restriction to each endpoint pair nonzero, so $F_P$ has weight exactly three.
Since $d\ge4$, $F_P$ is a stabilizer.
The map is injective because its restriction to $r$ is $P$, and its restriction to either $q_1$ or $q_2$ is also an isomorphism.
We therefore obtain a two-dimensional subspace $U_r\subset L_3(G)$.

Let $G_3$ count the chosen degree-$3$ partners, and let $z$ count the degree-$4$ partners assigned two isolated edges.
The star space, the $G_3$ one-dimensional subspaces, and the spaces $U_r$ form a direct sum.
Indeed, the star space vanishes on isolated edges, whereas every other summand has an injective restriction to isolated edges used by no other summand.
Hence,
\begin{equation}\label{eq:direct_sum_d4}
s+G_3+2z\le\dim L_3(G).
\end{equation}

Let $y$ count the degree-$4$ partners assigned one isolated edge.
A degree-$r\ge5$ column receives at most $\lfloor r/2\rfloor\le r-3$ isolated edges.
Since each degree-$4$ column contributes one to $t$,
\begin{equation}
M=G_3+y+2z+\sum_{r_q\ge5}\operatorname{occ}(q),\quad y+\sum_{r_q\ge5}\operatorname{occ}(q)\le t.
\end{equation}
Together with Eq.~\eqref{eq:direct_sum_d4}, this gives
\begin{equation}\label{eq:M_upper_d4}
M\le\dim L_3(G)-s+t.
\end{equation}
Combining this with Eq.~\eqref{eq:M_lower_bound} gives
\begin{equation}\label{eq:structure_d4}
m\le2\dim L_3(G)+s+b+2t.
\end{equation}

Combining Eqs.~\eqref{eq:e_identity}, \eqref{eq:I_degree_decomposition}, \eqref{eq:A_d4_definition}, and \eqref{eq:structure_d4}, we obtain
\begin{align}
3n&=I+e-t\label{eq:Delta_d4_identity}\\
&\le wm-(w-3)\dim L_3(G)-\dim L_2(G)+\frac{m-b+s}{2}-t\\
&\le wm-(w-4)\dim L_3(G)-\dim L_2(G)+s\\
&\le wm,
\end{align}
where the last inequality uses $w\ge4$ and $s\le\dim L_2(G)$.
This proves Eq.~\eqref{eq:d4_rate_weight_app} on the active core.
Restoring $h$ weight-$1$ factors increases $wm-3n$ by $(w-3)h\ge0$, so the same bound holds for the original code.
\end{proof}

Lemma~\ref{lem:wle3_distance} gives $w\ge4$ for $k\ge1$ and $d\ge3$.
Rearranging Eqs.~\eqref{eq:d3_rate_weight_app} and \eqref{eq:d4_rate_weight_app} and using the integrality of $w$ gives the general-code bounds in Theorem~\ref{thm:rate_weight_main} of the main text.

\subsection{CSS codes}

For CSS codes, we can choose pure $X$- and $Z$-generators without increasing the maximum weight.

\begin{lem}\label{lem:mixed_to_css_projection}
If $G$ is a CSS stabilizer group, then the optimal maximum generator weight is unchanged when the optimization is restricted to pure CSS generating sets; both optimizations give $\w(G)$.
\end{lem}

\begin{proof}
In binary notation, write $G=G_X\oplus G_Z$.
Suppose an arbitrary basis of $G$ consists of elements $g_i=(x_i,z_i)$ with $\wt(g_i)\le w$.
Then $(x_i,0)\in G_X$ and $(0,z_i)\in G_Z$, and each projection has weight at most $\wt(g_i)$.
The projections span $G$: for any pure $X$- or $Z$-stabilizer, projecting its expansion in the original basis gives an expansion in these projections.
Extracting an independent basis from the projections therefore gives a pure CSS basis of maximum weight at most $w$.
The reverse inequality is immediate because pure CSS bases form a subset of all Pauli bases.
\end{proof}

\begin{thm}\label{thm:css_rate_weight_app}
Let $G$ define a CSS $[\![n,k,d;w]\!]$ stabilizer code with $k\ge1$, $d\ge3$, and $w\ge4$.
Then
\begin{equation}\label{eq:css_refined_rate_weight_app}
4n\le(w+1)(n-k)-(w-3)\dim L_2(G).
\end{equation}
In particular,
\begin{equation}\label{eq:css_rate_weight_app}
\frac{k}{n}\le\frac{w-3}{w+1}.
\end{equation}
\end{thm}

\begin{proof}
First consider the active core.
Define $L_{2,X}\coloneqq L_2(G)\cap G_X$ and $L_{2,Z}\coloneqq L_2(G)\cap G_Z$.
Any mixed stabilizer of weight at most two has pure $X$ and $Z$ projections of weight at most two, so
\begin{equation}
L_2(G)=L_{2,X}\oplus L_{2,Z}.
\end{equation}
Let $\ell_X\coloneqq\dim L_{2,X}$, $\ell_Z\coloneqq\dim L_{2,Z}$, and $\ell\coloneqq\ell_X+\ell_Z=\dim L_2(G)$.
By Lemma~\ref{lem:mixed_to_css_projection}, we may choose a pure CSS basis of weight at most $w$, adapt it to pure bases of $L_{2,X}$ and $L_{2,Z}$, and obtain
\begin{equation}\label{eq:I_upper_css}
I\le2\ell+w(m-\ell)=wm-(w-2)\ell.
\end{equation}

Write $m=m_X+m_Z$, where $m_X$ and $m_Z$ are the numbers of $X$- and $Z$-rows, respectively, and let $a_q$ and $b_q$ be the corresponding numbers of rows incident on qubit $q$.
Lemma~\ref{lem:active_adapted_basis}(ii) gives $a_q,b_q\ge1$.
Define
\begin{equation}
s_X\coloneqq\#\{q:a_q=1\},\quad s_Z\coloneqq\#\{q:b_q=1\}.
\end{equation}
Fix an $X$-row $i$ and consider the columns on which it is the unique incident $X$-row.
The single-qubit $Z$ errors on these columns all have the same unit syndrome $\mathbf{e}_i$.
If at least one such column exists, their pairwise products span a space of weight-$2$ pure-$Z$ centralizer elements whose dimension is one less than the number of columns.
Since $d\ge3$, these elements are stabilizers.
The resulting spaces for distinct $X$-rows are independent.
There are at most $m_X$ such $X$-rows, so
\begin{equation}
s_X-m_X\le\ell_Z.
\end{equation}
Interchanging $X$ and $Z$ gives
\begin{equation}
s_Z-m_Z\le\ell_X.
\end{equation}
Therefore, $s_X+s_Z\le m+\ell$.
For each column,
\begin{equation}
a_q+b_q\ge4-\mathbf{1}_{a_q=1}-\mathbf{1}_{b_q=1},
\end{equation}
and hence
\begin{equation}\label{eq:I_lower_css}
I\ge4n-s_X-s_Z\ge4n-m-\ell.
\end{equation}
Combining Eqs.~\eqref{eq:I_upper_css} and \eqref{eq:I_lower_css} proves Eq.~\eqref{eq:css_refined_rate_weight_app} on the active core.

If the active core $G_0$ was obtained by removing $h$ weight-$1$ stabilizer factors, then $n=n_0+h$, $m=m_0+h$, and $\dim L_2(G)=\dim L_2(G_0)+h$.
Both sides of Eq.~\eqref{eq:css_refined_rate_weight_app} increase by exactly $4h$, so the refined inequality holds for the original code.
Dropping the nonnegative correction $(w-3)\dim L_2(G)$ gives Eq.~\eqref{eq:css_rate_weight_app}.
Rearranging $4n\le(w+1)(n-k)$ and using the integrality of $w$ gives $w\ge\left\lceil(3n+k)/(n-k)\right\rceil$.
Minimizing over CSS codes yields the corresponding check-weight bound in Theorem~\ref{thm:css_rate_weight_main} of the main text.
\end{proof}

The CSS bound is attained by surface codes constructed from complete graphs.

\begin{prop}\label{prop:tight_complete_graph_css}
For every integer $w\ge4$, there exists a CSS stabilizer code
\begin{equation}\label{eq:tight_complete_graph_css}
\left[\!\left[\frac{w(w+1)}2,\,\frac{w(w-3)}2,\,3;w\right]\!\right]
\end{equation}
that attains equality in Eq.~\eqref{eq:css_refined_rate_weight_app}, with $\dim L_2(G)=0$.
\end{prop}

\begin{proof}
Set $N\coloneqq w+1$, and take a self-dual cellular embedding of the complete graph $K_N$ in a closed surface.
An orientable self-dual embedding exists for $N=5$~\cite{naghipour2017classes}, and a nonorientable self-dual embedding exists for every $N\ge6$~\cite{archdeacon1994selfdual}.
The binary construction below applies to either type of surface.
Orientable complete-graph subfamilies and specific nonorientable instances have also appeared in the surface-code literature~\cite{naghipour2017classes,bhowmik2021surface}.

Place one qubit on each edge.
Let the $X$-checks be the vertex stars of the primal $K_N$, and let the $Z$-checks be the face boundaries, equivalently the vertex stars of the dual $K_N$.
The cellular boundary relation gives CSS orthogonality.
Since the primal and dual graphs are both connected, each incidence matrix has rank $N-1=w$ over $\mbb{F}_2$.
Therefore,
\begin{equation}
n=|E(K_N)|=\binom{N}{2}=\frac{w(w+1)}2,\quad m=2w,\quad k=n-m=\frac{w(w-3)}2.
\end{equation}
Both graphs are simple and $w$-regular, so every displayed check has weight $w$, and neither the primal nor the dual cycle space contains a nonzero vector of weight one or two.
The support of a mixed centralizer contains the support of each nonzero pure component, so the CSS distance is at least three.

The cycle space of $K_N$ is generated by triangles and has dimension
\begin{equation}
\binom{N}{2}-(N-1)=\frac{w(w-1)}2>w\quad(w\ge4),
\end{equation}
whereas the face-boundary stabilizer space has dimension $w$.
Thus at least one triangle is a weight-$3$ logical operator, giving $d=3$.

Finally, every nonzero $X$-stabilizer is a nonzero cut of the primal $K_N$ and hence has weight at least $N-1=w$; the same conclusion holds for $Z$-stabilizers in the dual graph.
The support of a mixed stabilizer contains the support of each nonzero component and therefore also has weight at least $w$.
Together with the weight-$w$ generating set, this gives $\w(G)=w$ and $\dim L_2(G)=0$.
Substitution yields
\begin{equation}
4n=(w+1)m,
\end{equation}
so the CSS bound is saturated.
Tensor products of arbitrarily many copies give an infinite tight family for each fixed $w$.
\end{proof}
\section{Weight enumerators and linear programming bounds}\label{app:enumerator_and_LP}

\begin{defn}[\cite{gottesman2016surviving}]
Let $\Pi$ be a projector from the $n$-qubit Hilbert space to a rank-$K$ subspace.
Define
\begin{align}
A_j&\coloneqq \frac{1}{K^2}\sum_{P\in\hat{\mc{P}}_n,\wt(P)=j}\abs{\Tr(P\Pi)}^2,\\
B_j&\coloneqq \frac{1}{K}\sum_{P\in\hat{\mc{P}}_n,\wt(P)=j}\Tr(P\Pi P\Pi),\\
Sh_j&\coloneqq \frac{1}{K}\sum_{P\in\hat{\mc{P}}_n,\wt(P)=j}\Tr(P\Pi PY^{\otimes n}\Pi^*Y^{\otimes n}),
\end{align}
where $\Pi^*$ is the complex conjugate of $\Pi$. We call $A_j$'s the \emph{weight enumerators} for $\Pi$.
Denote $\vec{A}=(A_0,A_1,\cdots,A_n)^{\mathrm{T}}$, $\vec{B}=(B_0,B_1,\cdots,B_n)^{\mathrm{T}}$, and $\vec{Sh}=(Sh_0,Sh_1,\cdots,Sh_n)^{\mathrm{T}}$.
\end{defn}

For any projector $\Pi$, we have~\cite{gottesman2016surviving}: i) $A_0=B_0=1$, ii) $B_j\ge A_j\ge0$, iii) $\Pi$ has distance $d$ iff $A_j=B_j$ for $j<d$, and iv) $Sh_j\ge0$.

Define $(n+1)\times(n+1)$ matrices $M$ and $\widetilde{M}$ by
\begin{equation}\label{eq:defn_of_M}
M_{ij}=P_i(j;n), \quad \widetilde{M}_{ij}=(-1)^jP_i(j;n),
\end{equation}
where $i,j=0,1,\cdots,n$, and $P_\cdot(\cdot,\cdot)$ is the Krawtchouk polynomial defined by
\begin{equation}
P_j(z;n)=\sum_{m=0}^{\min\{z,j\}}(-1)^m3^{j-m}\binom{n-z}{j-m}\binom{z}{m}.
\end{equation}

The quantum MacWilliams identity~\cite{Shor1997MacWilliams} states that
\begin{equation}
\vec{B}=\frac{K}{2^n}M\vec{A},
\end{equation}
and $\vec{Sh}$ can also be determined from $\vec{A}$ by~\cite{Rains1998Shadow,Rains1999shadow}
\begin{equation}
\vec{Sh}=\frac{K}{2^n}\widetilde{M}\vec{A}.
\end{equation}

Therefore, if there exists an $n$-qubit quantum code with logical dimension $K$ and distance $d$, then the following linear programming problem has a feasible solution:
\begin{subequations}\label{eq:LPbounds}
\begin{align}
\text{Find}\quad& A_0,A_1,\cdots,A_n\\
\text{s.t.}\quad& A_0=1,\,A_1\ge0,\cdots,A_n\ge0, \label{eq:LPbound_1}\\
&\frac{K}{2^n}M[:d]\vec{A}=\vec{A}[:d], \label{eq:LPbound_2}\\
&\frac{K}{2^n}M[d:]\vec{A}\ge\vec{A}[d:], \label{eq:LPbound_3}\\
&\widetilde{M}\vec{A}\ge0. \label{eq:LPbound_4}
\end{align}
\end{subequations}
Here $[:d]$ means taking the first $d$ rows (row 0 to row $d-1$) of a matrix, and $[d:]$ means taking the rest rows.

For an $[\![n,k]\!]$ stabilizer code with stabilizer group $G$, its weight enumerators satisfy~\cite{gottesman2016surviving}
\begin{equation}
A_j=\abs{\{g\in G: \wt(g)=j\}}\quad\text{for}\quad j=0,\cdots,n.
\end{equation}

An interesting observation is the following, which can be proved using the quantum MacWilliams identity~\cite{Shor1997MacWilliams}.
\begin{prop}
Suppose an $n$-qubit stabilizer code has $d\ge2$, then the average weight of the \emph{whole} stabilizer group is $\frac{1}{4}(3n-A_1)$, where $A_1$ is the number of weight-1 elements in its stabilizer group.
\end{prop}

\section{Linear constraints from weight}\label{app:linear_constraints_weight}

\subsection{Main constraints}

Suppose there exists an $[\![n,k,d;w]\!]$ stabilizer code with stabilizer group $G$.
Let $A_0,A_1,\cdots,A_n$ be the weight enumerators for this code.

We pick a weight-optimal generating set $\{g_1,\cdots,g_{n-k}\}$ for $G$, where $\wt(g_1)\le\cdots\le\wt(g_{n-k})$. By definition, we have $\wt(g_{n-k})=w$.

Let $y$ be the numbers of elements in $\{g_1,\cdots,g_{n-k}\}$ with weight $=w$, satisfying $1\le y\le n-k$.
We have $\wt(g_1)\le\cdots\le\wt(g_{n-k-y})<w=\wt(g_{n-k-y+1})=\cdots=\wt(g_{n-k})$.
Therefore, we have the constraint
\begin{equation}\label{eq:constraint1}
A_w\ge y.
\end{equation}
We now show that
\begin{equation}\label{eq:constraint2}
\sum_{i=0}^{w-1}A_i\le2^{n-k-y}.
\end{equation}
Suppose on the contrary that $\sum_{i=0}^{w-1}A_i>2^{n-k-y}$, then we can find a generating set $\{g_1'\cdots,g_{n-k}'\}$ of $G$ such that $\wt(g_1')\le\cdots\le\wt(g'_{n-k-y+1})<w$. However, notice that $\wt(g_{n-k-y+1})=w$, this contradicts the ``optimality" of $\{g_1,\cdots,g_{n-k}\}$.
As a special case, if $y=n-k$, \eqref{eq:constraint2} means $A_1=\cdots=A_{w-1}=0$.

For $p=0,...,n-k-y$ and $q=0,...,y$, an element in $G$ that can be written as the product of $p$ elements in $\{g_1,\cdots,g_{n-k-y}\}$ (weight $\le w-1$) and $q$ elements in $\{g_{n-k-y+1},\cdots,g_{n-k}\}$ (weight $=w$) has weight $\le p(w-1)+qw$.
Therefore, for every $(p,q)\in\{0,\cdots,n-k-y\}\times\{0,\cdots,y\}$, there exist $\binom{n-k-y}{p}\binom{y}{q}$ elements in $G$ that have weight $\le p(w-1)+qw$.
For any $M=w-1,\cdots,n$, we have
\begin{equation}\label{eq:constraint3}
\sum_{m=0}^MA_m\ge\sum_{p=0}^{n-k-y}\sum_{q=0}^{y}\mathbf{1}_{p(w-1)+qw\le M}\binom{n-k-y}{p}\binom{y}{q}.
\end{equation}
Here $\mathbf{1}$ denotes the indicator function.

Therefore, if an $[\![n,k,d;w]\!]$ stabilizer code exists, then there exists a $1\le y\le n-k$ such that \eqref{eq:constraint1}--\eqref{eq:constraint3} are satisfied.

Notice that, if for an $[\![n,k,d;w]\!]$ stabilizer code, $w$ satisfies $w=\lceil\frac{2n}{n-k}\rceil$, then by $d\ge2$ (see the proof of Theorem~\ref{thm:w_LBby_nk} in the main text) we must have $wy+(w-1)(n-k-y)\ge2n$, implying $y\ge2n-(w-1)(n-k)$.

\subsection{Stabilizer code constraint}

For an $[\![n,k]\!]$ stabilizer code, write $G=\langle S\rangle$ with $\abs{S}=n-k$. If all elements in $S$ have even weights, then all elements in $G$ have even weights; if there exists an element in $S$ with odd weight, then half of the elements in $G$ have even weights.
Therefore, for an $[\![n,k]\!]$ stabilizer code, one has
\begin{equation}
\text{either}\quad\sum_{i=0}^{\lfloor n/2\rfloor}A_{2i}=2^{n-k-1}\quad\text{or}\quad\sum_{i=0}^{\lfloor n/2\rfloor}A_{2i}=2^{n-k}.
\end{equation}

Furthermore, if for an $[\![n,k,d;w]\!]$ stabilizer code, $w$ is odd, then
\begin{equation}
\sum_{i=0}^{\lfloor n/2\rfloor}A_{2i}=2^{n-k-1}.
\end{equation}
If for an $[\![n,k,d;w]\!]$ stabilizer code, $w$ is even, and $y=n-k$ (where $y$ is the number of elements in a weight-optimal generating set that has weight $w$), then
\begin{equation}
\sum_{i=0}^{\lfloor n/2\rfloor}A_{2i}=2^{n-k}.
\end{equation}

\subsection{$A_1=0$ constraint}

Suppose there exists an $[\![n,k,d;w]\!]$ stabilizer code with stabilizer group $G$ and enumerator $A_1>0$.
By Lemma~\ref{lemma:clean_weight1}, there exists an $[\![n-A_1,k,d;w]\!]$ stabilizer code, thus there also exists an $[\![n-1,k,d;w]\!]$ stabilizer code.

Therefore, suppose $w<\mathrm{W}_{\!\mathrm{opt}}(n-1,k,d)$ and there exists an $[\![n,k,d;w]\!]$ stabilizer code, then we must have the constraint
\begin{equation}
A_1=0.
\end{equation}
In practice, $w<\mathrm{W}_{\!\mathrm{opt}}(n-1,k,d)$ can be inferred from $w<\mathrm{W}_{\!\mathrm{LB}}(n-1,k,d)$, where $\mathrm{W}_{\!\mathrm{LB}}\le \mathrm{W}_{\!\mathrm{opt}}$ is an efficiently computable lower bound.

\subsection{Constraint imposed by generators' overlap}

If two commuting Pauli strings $P$ and $Q$ have overlap in their supports, then we have $\wt(PQ)\le\wt(P)+\wt(Q)-2$.

Suppose $G$ is an $[\![n,k,d;w]\!]$ stabilizer code that is not a simple product of smaller codes.
we pick a generating set $\{g_1,\cdots,g_{n-k}\}$ for $G$, with $\wt(g_1)\le\cdots\le\wt(g_{n-k})=w$.
Form the \textit{overlap graph} on these generators: the $n-k$ vertices correspond to the $g_i$’s, and we place an edge between $g_i$ and $g_j$ iff their supports overlap.
This graph must be connected; otherwise the generators split into two disjoint sets, and the code would factor.
A connected graph on $n-k$ vertices has at least $n-k-1$ edges.
Therefore, for all the $\binom{n-k}{2}$ pairs of generators, at least $n-k-1$ pairs are overlapped, so their product satisfies $1\le$ weight $\le 2w-2$.
Note that the $n-k$ generators satisfies $1\le$ weight $\le w\le 2w-2$.
Consequently, we have the constraint
\begin{equation}
\sum_{i=1}^{\min\{2w-2,n\}}A_i\ge (n-k)+(n-k-1).
\end{equation}
Notice that the number $n-k-1$ cannot be improved without further information about the code, since for a $[\![4,2,2;4]\!]$ code, there are $n-k=2$ generators and $n-k-1=1$ edge.

We give a necessary condition for an $[\![n,k,d;w]\!]$ stabilizer code to factorize into smaller codes.
Suppose an $[\![n,k,d;w]\!]$ code can be written as a tensor product of $T\ge2$ codes with parameters
$[\![n_i,k_i,d_i;w_i]\!]$, $i=1,\cdots,T$, where $1\le n_i\le n-1$, and
\begin{equation}
n=n_1+\cdots+n_T,\quad k=k_1+\cdots+k_T.
\end{equation}
By Lemma~\ref{lemma:tensor_product_parameters}, we know $d_i\ge d$, $w_i\le w$.
Then, there exists an $i$ such that the code $[\![n_i,k_i,d_i;w_i]\!]$ has rate $k_i/n_i\ge k/n$.
Therefore, there must exist a stabilizer code with parameters $[\![n',k',d';w']\!]$ with $1\le n'\le n-1$, $k'/n'\ge k/n$, $d'\ge d$, $w'\le w$.

If all codes with size $1\le n'\le n-1$ with distance $d'\ge d$ and weight $w'\le w$ have code rate $k'/n'< k/n$, then we know that any $[\![n,k,d;w]\!]$ code, if exists, cannot be factorized into tensor products of smaller codes.
Suppose $\mathrm{W}_{\!\mathrm{LB}}\le \mathrm{W}_{\!\mathrm{opt}}$ is an efficiently computable lower bound for $\mathrm{W}_{\!\mathrm{opt}}$. Then
\begin{align}
&\begin{aligned}
&\big\{(n',k'): \; 1\le n'\le n-1,\ \exists\, d'\ge d,\\
&\quad w'\le w,\ \text{and an }[\![n',k',d';w']\!]\ \text{code}\big\}\label{eq:setA}
\end{aligned}\\
&\begin{aligned}
=&\big\{(n',k'): \; 1\le n'\le n-1,\ \exists\, d'\ge d\\
&\quad\text{such that}\ \mathrm{W}_{\!\mathrm{opt}}(n',k',d')\le w\big\}\label{eq:setB}
\end{aligned}\\
&\begin{aligned}
\subset&\big\{(n',k'): \; 1\le n'\le n-1,\ \exists\, d'\ge d\\
&\quad\text{such that}\ \mathrm{W}_{\!\mathrm{LB}}(n',k',d')\le w\big\}
=:R_{d,w}(n).\label{eq:setC}
\end{aligned}
\end{align}
Equality \eqref{eq:setA}=\eqref{eq:setB} holds by the definition of $\mathrm{W}_{\!\mathrm{opt}}$ (recall $\mathrm{W}_{\!\mathrm{opt}}(n',k',d')=\infty$ if no
$[\![n',k',d']\!]$ code exists).
Consequently, if
\begin{equation}
r_{d,w}(n)\coloneqq \max_{(n',k')\in R_{d,w}(n)}k'/n'<k/n,
\end{equation}
then any $[\![n,k,d;w]\!]$ code (if it exists) cannot be factorized into tensor products of smaller codes.
Notice that $r_{d,w}(n)$ is efficiently computable.

\subsection{Summary}

We summarize the linear constraints in the above subsections.

Assume there is an $[\![n,k,d;w]\!]$ stabilizer code. Then there exists an admissible choice of
\begin{align}
&y\in \big\{\max\{1,2n-(w-1)(n-k)\},\cdots,n-k\big\},\\
&\mathrm{parity}\in\{0,1\},\\
&b_{\mathrm{single}},\,b_{\mathrm{overlap}}\in\{0,1\},
\end{align}
such that the following linear program $\mathrm{LP}(n,k,d;w;y,\mathrm{parity},b_{\mathrm{single}},b_{\mathrm{overlap}})$ is feasible in the variables $A_0,\cdots,A_n$:
\begin{subequations}\label{eq:LPbounds_weight}
\begin{align}
\text{Find}\quad& A_0,A_1,\cdots,A_n\in\mbb{R}\\
\text{s.t.}\quad&\text{ Constraints~\eqref{eq:LPbound_1}-\eqref{eq:LPbound_4} hold,}\\
&A_w\ge y,\\
&A_0+\cdots+A_{w-1}\le2^{n-k-y},\\
&\begin{aligned}
&\sum_{m=0}^MA_m\ge\sum_{p=0}^{n-k-y}\sum_{q=0}^{y}\mathbf{1}_{p(w-1)+qw\le M}\times\\
&\;\binom{n-k-y}{p}\binom{y}{q},\text{ for }M=w-1,\cdots,n-1,\\
\end{aligned}\\
&\sum_{i=0}^{\lfloor n/2\rfloor}A_{2i}=2^{n-k-\mathrm{parity}},\\
&b_{\text{single}}\cdot A_1=0,\\
&b_{\text{overlap}}\cdot\Big[\sum_{i=1}^{\min\{2w-2,n\}}A_i-(2n-2k-1)\Big]\ge0.
\end{align}
\end{subequations}
Constraints \eqref{eq:LPbound_1}-\eqref{eq:LPbound_4} are those that must hold if there exists an $[\![n,k,d]\!]$ code.
The admissible choices of $\mathrm{parity}$, $b_{\mathrm{single}}$, and $b_{\mathrm{overlap}}$ are limited for certain parameter candidate $[\![n,k,d;w]\!]$, given by:
\begin{equation}
\mathrm{parity}=
\begin{cases}
1, & \text{if } w \text{ is odd},\\
0, & \text{if } w \text{ is even and } y=n-k,\\
\text{either }0\text{ or }1, & \text{otherwise},
\end{cases}
\end{equation}
and
\begin{align}
b_{\mathrm{single}}&=\mathbf{1}_{w<\mathrm{W}_{\!\mathrm{LB}}(n-1,k,d)},\\
b_{\mathrm{overlap}}&=\mathbf{1}_{r_{d,w}(n)<k/n}.
\end{align}
If this LP is infeasible for all admissible choices of
$y$, $\mathrm{parity}$, $b_{\mathrm{single}}$, and $b_{\mathrm{overlap}}$, then no $[\![n,k,d;w]\!]$ stabilizer code exists.

\begin{thm}\label{thm:WLB_LP_certificate_app}
Fix integers $(n,k,d)$ and a candidate generator weight $w$.
Let $\mathrm{LP}(n,k,d;w;y,\mathrm{parity},b_{\mathrm{single}},b_{\mathrm{overlap}})$ denote the feasibility problem
\eqref{eq:LPbounds_weight} in the variables $A_0,\cdots,A_n$, where the discrete parameters are required to satisfy
\begin{equation}
y\in \{\max\{1,2n-(w-1)(n-k)\},\cdots,n-k\},\quad
\mathrm{parity}\in\{0,1\},\quad
b_{\mathrm{single}},b_{\mathrm{overlap}}\in\{0,1\},
\end{equation}
together with the admissibility rules for $\mathrm{parity}$ and the definitions of $b_{\mathrm{single}}=\mathbf{1}_{w<\mathrm{W}_{\!\mathrm{LB}}(n-1,k,d)}$ and $b_{\mathrm{overlap}}=\mathbf{1}_{r_{d,w}(n)<k/n}$.
If for \emph{every} admissible choice of $(y,\mathrm{parity},b_{\mathrm{single}},b_{\mathrm{overlap}})$ the problem $\mathrm{LP}(n,k,d;w;y,\mathrm{parity},b_{\mathrm{single}},b_{\mathrm{overlap}})$ is infeasible, then there is no $[\![n,k,d;w]\!]$ stabilizer code.
\end{thm}

\section{Finite-length lower bounds in the rate--distance plane}\label{app:rate_distance}

\begin{figure}
    \centering
    \includegraphics[width=0.9\textwidth]{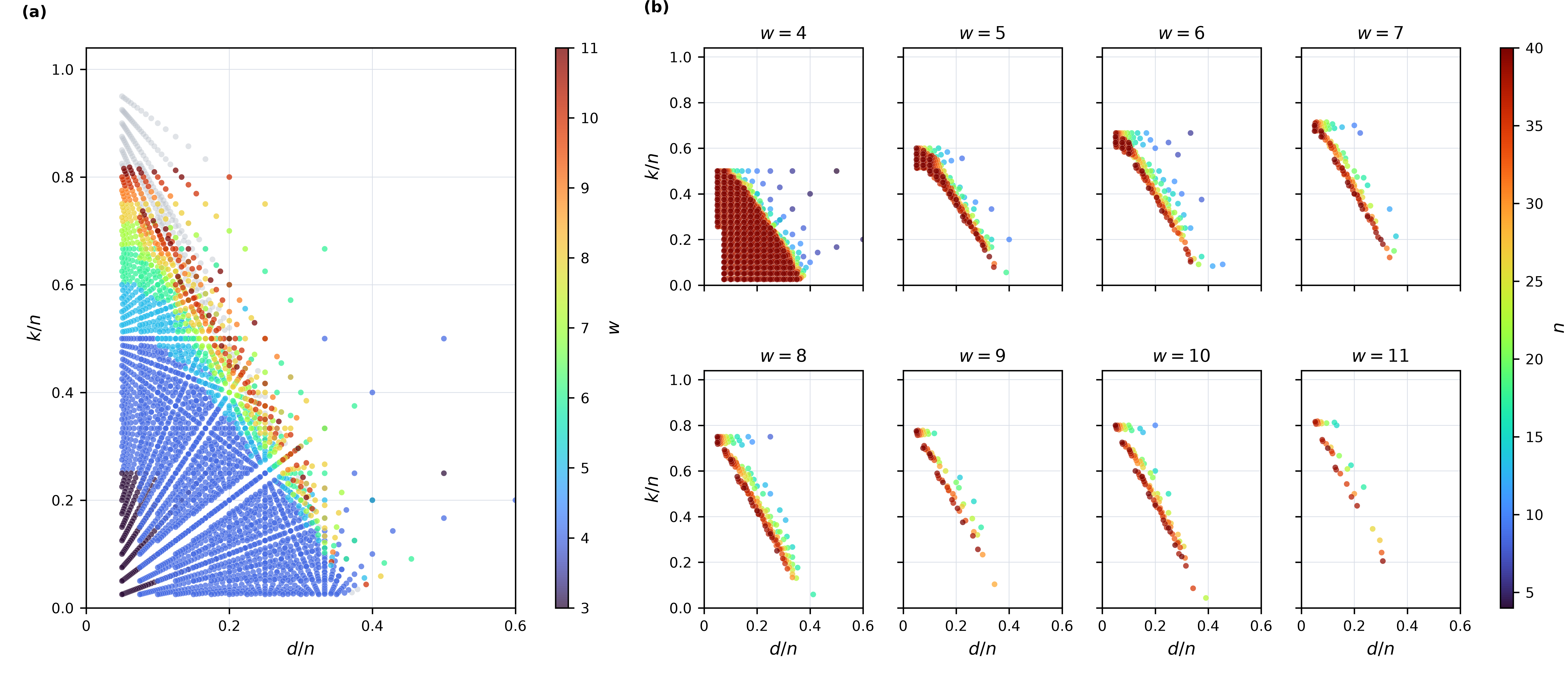}
    \caption{Code rate $k/n$ versus relative distance $d/n$ for the finite-length lower-bound data. (a) All points, colored by the lower-bound value $w\leq 11$, with points having $w>11$ shown in gray. (b) The subset with $4\leq w\leq 11$, separated by $w$ and colored by block length $n$.}
    \label{fig:ratevsd}
\end{figure}

Fig.~\ref{fig:ratevsd} maps the finite-length lower bound $\wlb(n,k,d)$ in the rate--relative-distance plane. The upper edge of each $w$ band follows the rate ceiling $k/n\leq 1-2/w$ from Theorem~\ref{thm:w_LBby_nk} in the main text, while its downward slope with $d/n$ shows the additional restriction imposed by distance. In panel (b), the larger-$n$ points trace these bands more sharply, whereas the scattered small-$n$ points reveal stronger finite-size and integrality effects.

\section{Low-weight code constructions}\label{app:Low-weight_code_constructions}

\begingroup
\small
\noindent$[\![4,1,2;3]\!]$: $\langle XXXI,IYYY,ZIZZ\rangle$\\
$[\![4,2,2;4]\!]$: $\langle XXXX,ZZZZ\rangle$\\
$[\![5,1,2;3]\!]$: $[\![4,1,2;3]\!]\otimes[\![1,0]\!]$\\
$[\![5,2,2;4]\!]$: $[\![4,2,2;4]\!]\otimes[\![1,0]\!]$\\
$[\![5,1,3;4]\!]$: $\langle XZZXI,IXZZX,XIXZZ,ZXIXZ\rangle$\\
$[\![6,1,2;3]\!]$: $[\![4,1,2;3]\!]\otimes[\![2,0]\!]$\\
$[\![6,2,2;4]\!]$: $[\![4,2,2;4]\!]\otimes[\![2,0]\!]$\\
$[\![6,3,2;4]\!]$: $\langle XXXXII,IIYYYY,ZZIIZZ\rangle$\\
$[\![6,4,2;6]\!]$: $\langle XXXXXX,ZZZZZZ\rangle$\\
$[\![6,1,3;4]\!]$: $[\![5,1,3;4]\!]\otimes[\![1,0]\!]$\\
$[\![7,1,2;3]\!]$: $[\![4,1,2;3]\!]\otimes[\![3,0]\!]$\\
$[\![7,2,2;4]\!]$: $[\![4,2,2;4]\!]\otimes[\![3,0]\!]$\\
$[\![7,3,2;4]\!]$: $[\![6,3,2;4]\!]\otimes[\![1,0]\!]$\\
$[\![7,4,2;6]\!]$: $[\![6,4,2;6]\!]\otimes[\![1,0]\!]$\\
$[\![7,1,3;4]\!]$: $[\![5,1,3;4]\!]\otimes[\![2,0]\!]$\\
$[\![8,1,2;3]\!]$: $[\![4,1,2;3]\!]\otimes[\![4,0]\!]$\\
$[\![8,2,2;3]\!]$: $[\![4,1,2;3]\!]^{\otimes 2}$\\
$[\![8,3,2;4]\!]$: $[\![6,3,2;4]\!]\otimes[\![2,0]\!]$\\
$[\![8,4,2;4]\!]$: $[\![4,2,2;4]\!]^{\otimes 2}$\\
$[\![8,5,2;6]\!]$: $\langle XXXXXXII,IIYYYYYY,ZZIIZZZZ\rangle$\\
$[\![8,6,2;8]\!]$: $\langle XXXXXXXX,ZZZZZZZZ\rangle$\\
$[\![8,1,3;4]\!]$: $[\![5,1,3;4]\!]\otimes[\![3,0]\!]$\\
$[\![8,2,3;4]\!]$:
$\langle ZIIIIYZY, IIIZXIXZ, ZZIYYIII, IZXIIXIZ, IXZIZIZI, XIZZIZII\rangle$~\cite{cross2025small}\\
$[\![8,3,3;6]\!]$: $\langle XIZIYZXY, ZIZXIYYX, IXZZYXYI, IZIYZXXY, IIXYXZYZ\rangle$~\cite{grasslCodetables}\\
$[\![9,1,2;3]\!]$: $[\![4,1,2;3]\!]\otimes[\![5,0]\!]$\\
$[\![9,2,2;3]\!]$: $[\![4,1,2;3]\!]^{\otimes 2}\otimes[\![1,0]\!]$\\
$[\![9,3,2;4]\!]$: $[\![6,3,2;4]\!]\otimes[\![3,0]\!]$\\
$[\![9,4,2;4]\!]$: $[\![4,2,2;4]\!]^{\otimes 2}\otimes[\![1,0]\!]$\\
$[\![9,5,2;5]\!]$: $\langle IIIIXXXXX, IIZZXYYII, ZZIXZIIIZ, YYXIZIIZI\rangle$~\cite{cross2025small}\\
$[\![9,6,2;7]\!]$: $\langle IIIXXXXXX, XXXIIZZZZ, ZZZZZIIZY\rangle$~\cite{cross2025small}\\
$[\![9,1,3;4]\!]$: $[\![5,1,3;4]\!]\otimes[\![4,0]\!]$\\
$[\![9,2,3;4]\!]$: $[\![8,2,3;4]\!]\otimes[\![1,0]\!]$\\
$[\![9,3,3;5]\!]$: $\langle IIIIXXXXX, IXIZYIZXI, IYYIIXIZY, XZXIXXIII, YIIZIYIXY,
 ZIYYIIYIY\rangle$~\cite{cross2025small}\\
$[\![10,1,2;3]\!]$: $[\![4,1,2;3]\!]\otimes[\![6,0]\!]$\\
$[\![10,2,2;3]\!]$: $[\![4,1,2;3]\!]^{\otimes 2}\otimes[\![2,0]\!]$\\
$[\![10,3,2;4]\!]$: $[\![6,3,2;4]\!]\otimes[\![4,0]\!]$\\
$[\![10,4,2;4]\!]$: $[\![4,2,2;4]\!]^{\otimes 2}\otimes[\![2,0]\!]$\\
$[\![10,5,2;4]\!]$: $[\![6,3,2;4]\!]\otimes[\![4,2,2;4]\!]$\\
$[\![10,6,2;6]\!]$: $\langle XXXXXXIIII, ZZZZZZIIII, IIIIXXXXXX,IIIIZZZZZZ\rangle$\\
$[\![10,7,2;7]\!]$: $\langle XXXXXXXIII,IIIYYYYYYY,ZZZIIIZZZZ\rangle$\\
$[\![10,8,2;10]\!]$: $\langle XXXXXXXXXX, ZZZZZZZZZZ\rangle$\\
$[\![10,1,3;4]\!]$: $[\![5,1,3;4]\!]\otimes[\![5,0]\!]$\\
$[\![10,2,3;4]\!]$: $[\![8,2,3;4]\!]\otimes[\![2,0]\!]$\\
$[\![10,3,3;5]\!]$: $[\![9,3,3;5]\!]\otimes[\![1,0]\!]$\\
$[\![10,4,3;7]\!]$ (not known to be optimal):~\cite[Fig.~11]{Roffe2020Graphical}\\
$[\![10,1,4;6]\!]$ (not known to be optimal): The below $[\![10,2,4;6]\!]$ code plus a weight 4 logical Pauli.\\
$[\![10,2,4;6]\!]$ (not known to be optimal): $\langle XIIIXIYZZY, ZIIIXYIYXZ, IXIIYZYIZX, IZIIIZZZYY, IIXIIXZXXZ,\allowbreak IIZIZIYYXX, IIIXYZZYXI, IIIZXYXZYI\rangle$~\cite{grasslCodetables}\\
$[\![11,1,2;3]\!]$: $[\![4,1,2;3]\!]\otimes[\![7,0]\!]$\\
$[\![11,2,2;3]\!]$: $[\![4,1,2;3]\!]^{\otimes 2}\otimes[\![3,0]\!]$\\
$[\![11,3,2;4]\!]$: $[\![6,3,2;4]\!]\otimes[\![5,0]\!]$\\
$[\![11,4,2;4]\!]$: $[\![4,2,2;4]\!]^{\otimes 2}\otimes[\![3,0]\!]$\\
$[\![11,5,2;4]\!]$: $[\![6,3,2;4]\!]\otimes[\![4,2,2;4]\!]\otimes[\![1,0]\!]$\\
$[\![11,6,2;5]\!]$: $\langle IIIIIIXXXXI,IIIIIXZZZZI,IIIXXZIIIXX,XXXIZIIIIIZ,ZZZZIIIIIIY\rangle$\\
$[\![11,7,2;6]\!]$: $\langle XXXXIIIIIXX,IIIIXXXXXIX,ZZIIIZIIIZZ,IIZZZIZZZII\rangle$\\
$[\![11,8,2;8]\!]$: $\langle IIIIXXXXXXZ,XXXXZZZIIIX,ZZZZIIIZZZX\rangle$\\
$[\![11,1,3;4]\!]$: $[\![5,1,3;4]\!]\otimes[\![6,0]\!]$\\
$[\![11,2,3;4]\!]$: $[\![8,2,3;4]\!]\otimes[\![3,0]\!]$\\
$[\![11,3,3;5]\!]$ (not known to be optimal): $[\![9,3,3;5]\!]\otimes[\![2,0]\!]$\\
$[\![11,4,3;7]\!]$ (not known to be optimal): $[\![10,4,3;7]\!]\otimes[\![1,0]\!]$\\
$[\![11,5,3;8]\!]$ (not known to be optimal): $\langle XIYIXYXXYZI,ZIXIZXZZXYI,IXXIIYYZZZZ,IZZIIXXYYYY,\allowbreak IIIXXZZYYYY,IIIZZYYXXXX \rangle$~\cite{grasslCodetables}\\
$[\![11,1,4;6]\!]$ (not known to be optimal): $[\![10,1,4;6]\!]\otimes[\![1,0]\!]$\\
$[\![11,2,4;6]\!]$ (not known to be optimal): $[\![10,2,4;6]\!]\otimes[\![1,0]\!]$\\
$[\![11,1,5;6]\!]$: $\langle XIIIIXZZIXX, ZIIIIZXYYIX, IXIIIXZYZIY,IZIIIZZZYYI, IIXIIXZXXZI,IIZIIZYIYZZ,\allowbreak IIIXIXYIXXY,IIIZIZIZXZX,IIIIXXXIZZX,IIIIZZIYZYZ\rangle$~\cite{grasslCodetables}\\
$[\![12,1,2;3]\!]$: $[\![4,1,2;3]\!]\otimes[\![8,0]\!]$\\
$[\![12,2,2;3]\!]$: $[\![4,1,2;3]\!]^{\otimes 2}\otimes[\![4,0]\!]$\\
$[\![12,3,2;3]\!]$: $[\![4,1,2;3]\!]^{\otimes 3}$\\
$[\![12,4,2;4]\!]$: $[\![4,2,2;4]\!]^{\otimes 2}\otimes[\![4,0]\!]$\\
$[\![12,5,2;4]\!]$: $[\![6,3,2;4]\!]\otimes[\![4,2,2;4]\!]\otimes[\![2,0]\!]$\\
$[\![12,6,2;4]\!]$: $[\![6,3,2;4]\!]^{\otimes2}$\\
$[\![12,7,2;6]\!]$: $[\![11,7,2;6]\!]\otimes[\![1,0]\!]$\\
$[\![12,8,2;6]\!]$: $\langle XXXXIIIIIIXX,IIIIXXXXXXII,ZZIIZZIIZZII,IIZZIIZZIIZZ\rangle$\\
$[\![12,9,2;8]\!]$:
$\langle XXXXXXXXIIII,IIIIYYYYYYYY,ZZZZIIIIZZZZ\rangle$\\
$[\![12,1,3;4]\!]$: $[\![5,1,3;4]\!]\otimes[\![7,0]\!]$\\
$[\![12,2,3;4]\!]$: $[\![8,2,3;4]\!]\otimes[\![4,0]\!]$\\
$[\![12,3,3;5]\!]$ (not known to be optimal): $[\![9,3,3;5]\!]\otimes[\![3,0]\!]$\\
$[\![12,4,3;7]\!]$ (not known to be optimal): $[\![10,4,3;7]\!]\otimes[\![2,0]\!]$\\
$[\![12,5,3;8]\!]$ (not known to be optimal): $[\![11,5,3;8]\!]\otimes[\![1,0]\!]$\\
$[\![12,6,3;10]\!]$ (not known to be optimal): $\langle XIIYIXIZYXYX,ZIIXIZIYXZXZ,IXIYZYYXIZZI,IZIXYXXZIYYI,\allowbreak IIXXXXXXXXXX,IIZZZZZZZZZZ\rangle$~\cite{grasslCodetables}\\
$[\![12,1,4;6]\!]$ (not known to be optimal): $[\![10,1,4;6]\!]\otimes[\![2,0]\!]$\\
$[\![12,2,4;6]\!]$ (not known to be optimal): $[\![10,2,4;6]\!]\otimes[\![2,0]\!]$\\
$[\![12,3,4;8]\!]$ (not known to be optimal): The below $[\![12,4,4;8]\!]$ code plus a weight 4 logical Pauli.\\
$[\![12,4,4;8]\!]$ (not known to be optimal): $\langle XIIXYZIXZIZX,ZIIZXYIZYIYZ,IXIXZYIXIZXZ,IZIZYXIZIYZY,\allowbreak IIXXXXIIXXZZ, IIZZZZIIZZYY,IIIIIIXXXXXX,IIIIIIZZZZZZ\rangle$~\cite{grasslCodetables}\\
$[\![12,1,5;6]\!]$: $[\![11,1,5;6]\!]\otimes[\![1,0]\!]$\\
\endgroup

\section{Lower bound for specific parameters}\label{app:specific_parameter_bound}

By Algorithm~\ref{algo:Weight_LB_Table}, we can get $\wopt(12,7,2)\ge5$.
\begin{prop}\label{prop:LB_12_7_2}
$\wopt(12,7,2)\ge6$.
\end{prop}
\begin{proof}
Assume for contradiction that there exist $5$ independent commuting generators
$g_1,\cdots,g_5\in\mc P_{12}^+$ with $\wt(g_i)\le 5$ and that the resulting code has distance $d\ge 2$.

For each qubit $j\in[12]$, let
\begin{equation}
t_j \coloneqq \#\{ i\in[5] : \text{$g_i$ acts nontrivially on qubit $j$}\}.
\end{equation}
Since $\wopt(11,7,2)=6>5$, we know $t_j\ge2$ for all $j$. We have
\begin{equation}
\label{eq:sum_tj}
24\le\sum_{j=1}^{12} t_j = \sum_{i=1}^{5}\wt(g_i)\le 25.
\end{equation}

If $\sum_j t_j=24$, then $t_j=2$ for all $j$. For different $a,b\in[5]$, when $\supp(g_a)$ and $\supp(g_b)$ overlap, they must differ on every qubit they overlap (otherwise $d=1$).
Since any pair $g_a,g_b$ commute, we know
\begin{equation}
\label{eq:even_overlap_caseA}
|\supp(g_a)\cap\supp(g_b)| \equiv 0 \pmod 2
\quad\text{for all }a\ne b.
\end{equation}
Fix $a$, since each qubit in $\supp(g_a)$ is shared with exactly one other generator, we have
\begin{equation}
\wt(g_a)=\sum_{b\ne a} |\supp(g_a)\cap\supp(g_b)|.
\end{equation}
By \eqref{eq:even_overlap_caseA}, the right-hand side is a sum of even integers, hence $\wt(g_a)$ is even.
But $\wt(g_a)\le 5$, so $\wt(g_a)\in\{2,4\}$ for all $a$, which is impossible.

If $\sum_j t_j=25$, then exactly one qubit has $t_j=3$ and all other qubits have $t_j=2$. Also, $\wt(g_i)=5$ for all $i\in[5]$.
Let $j^\star$ be the unique qubit with $t_{j^\star}=3$. Exactly three generators act nontrivially on $j^\star$,
so there exists at least one generator, say $g_5$, that acts trivially on $j^\star$.

Every qubit $j$ in $\supp(g_5)$ is not $j^\star$, hence has $t_j=2$,
so it is shared by $g_5$ and exactly one other generator. By $[g_5,g_b]=0$, we know
\begin{equation}
\label{eq:even_overlap_caseB}
|\supp(g_5)\cap\supp(g_b)| \equiv 0 \pmod 2
\quad \forall b\in\{1,2,3,4\}.
\end{equation}
Again we have
\begin{equation}
\wt(g_5)=\sum_{b=1}^4 |\supp(g_5)\cap\supp(g_b)|,
\end{equation}
so by \eqref{eq:even_overlap_caseB}, $\wt(g_5)$ is even, contradicting $\wt(g_5)=5$.
\end{proof}

\clearpage
\section{Pseudo code for calculating weight lower bound}\label{app:Pseudo_code}

We summarize the procedure of calculating a lower bound $\mathrm{W}_{\!\mathrm{LB}}$ for $\mathrm{W}_{\!\mathrm{opt}}$, for all stabilizer codes with $n\le N$, into Algorithm~\ref{algo:Weight_LB_Table}.

\begin{algorithm}[H]
\caption{Calculating $\mathrm{W}_{\!\mathrm{LB}}$}\label{algo:Weight_LB_Table}
\begin{algorithmic}[1]
\Require
An integer $N$.
\Ensure
$\mathrm{W}_{\!\mathrm{LB}}(n,k,d)$ for $4\le n\le N$, $2\le d\le n$, $2\le k\le n$.
\For{$n = 4, \cdots, N$}
\For{$d = 2, \cdots, \lfloor\frac{n+1}{2}\rfloor$}\Comment{By quantum Singleton bound $n-k\ge2(d-1)$.}
\For{$k = 1, \cdots, n$}
\If{$d = 2$ and $k/n \le 1/4$}
\State Set $\mathrm{W}_{\!\mathrm{LB}}(n,k,d) = 3$ \Comment{Based on Theorem~\ref{thm:w_LBby_nk} in the main text.}
\State \textbf{continue}
\EndIf
\If{LP problem \eqref{eq:LPbounds} is infeasible}
\For{$\kappa = k, \cdots, n$}
\State Set $\mathrm{W}_{\!\mathrm{LB}}(n, \kappa, d) = \infty$
\EndFor
\State \textbf{break}
\EndIf
\State Set $\mathrm{BreakFlag} = \text{False}$
\For{$w = \max\{4, \lceil \frac{2n}{n-k} \rceil\}, \cdots, n$} \Comment{Based on Theorem~\ref{thm:wle3} and Theorem~\ref{thm:w_LBby_nk} in the main text.}
\State Set $b_{\mathrm{single}} = \mathbf{1}_{\,w < \mathrm{W}_{\!\mathrm{LB}}(n-1, k, d)}$
\State Set $b_{\mathrm{overlap}} = \mathbf{1}_{\,r_{d,w}(n) < k/n}$
\For{$y = \max\{1, 2n - (w-1)(n-k)\}, \cdots, n-k$}
\For{$\mathrm{parity} = 0, 1$}
\If{($\mathrm{parity} = 0$ and $w$ is odd) or ($\mathrm{parity} = 1$ and $w$ is even and $y = n-k$)}
\State \textbf{continue}
\EndIf
\If{$\mathrm{LP}(n,k,d;w;y,\mathrm{parity},b_{\mathrm{single}},b_{\mathrm{overlap}})$~\eqref{eq:LPbounds_weight} is feasible}
\State Set $\mathrm{W}_{\!\mathrm{LB}}(n, k, d) = w$
\State Set $\mathrm{BreakFlag} = \text{True}$
\State \textbf{break}
\EndIf
\EndFor
\If{$\mathrm{BreakFlag} = \text{True}$}
\State \textbf{break}
\EndIf
\EndFor
\If{$\mathrm{BreakFlag} = \text{True}$}
\State \textbf{break}
\EndIf
\EndFor
\If{$\mathrm{BreakFlag} = \text{False}$}
\State Set $\mathrm{W}_{\!\mathrm{LB}}(n, k, d) = \infty$
\EndIf
\EndFor
\EndFor
\EndFor
\end{algorithmic}
\end{algorithm}

\end{document}